\documentclass{article}
\RequirePackage{natbib}
\usepackage{amsmath}
\usepackage{upgreek}
\usepackage{graphicx}
\usepackage{amssymb}
\providecommand\ttau{\tilde\tau}
\providecommand\tx{\tilde{x}}
\providecommand\ty{\tilde{y}}
\providecommand\dd{\mathrm{d}}
\providecommand\Ja{\mbox{\textit{Ja}}}
\providecommand\Pen{\mbox{\textit{Pe}}}
\providecommand\tT{\tilde T}
\providecommand\cgrad{\left.\frac{\partial c}{\partial \xi}\right|_{\xi = 1}}
\providecommand\acorr{a_\mathit{corr}}
\providecommand\deltacorr{\delta_\mathit{corr}}
\providecommand\upi{\uppi}
\providecommand\umu{\upmu}
\newcommand\Real{\mbox{Re}}          % cf plain TeX's \Re and Reynolds number
\newcommand\Imag{\mbox{Im}}           % cf plain TeX's \Im

\begin{document}

%\newtheorem{lemma}{Lemma}
%\newtheorem{corollary}{Corollary}
%
%\shorttitle{The history effect in bubble growth and dissolution. Part 1.} 
%%for header on odd pages
%\shortauthor{P. Pe\~nas-L\'opez et al} 
%%for header on even pages
%
\title{The history effect in bubble growth and dissolution. Part 1. Theory}
\author{
 Pablo Pe\~nas-L\'opez$^1$,
 Miguel A. Parrales$^1$, \\
 Javier Rodr\'iguez-Rodr\'iguez$^1$
 and
 Devaraj van der Meer$^2$}

\maketitle
\setcounter{footnote}{-1}
\footnote{$^1$
Fluid Mechanics Group, Universidad Carlos III de Madrid, Avda. de la Universidad 30, 28911 Legan\'es (Madrid), Spain}
\setcounter{footnote}{-1}
\footnote{$^2$
Physics of Fluids Group, Faculty of Science and Technology, University of Twente, P.O. Box 217, 7500 AE Enschede, The Netherlands
}

%%%%%%%%%%%%%%%%%%%%%%%%%%%%%%%%%%%%%%%%%%%%%%
%% Abstract
%%%%%%%%%%%%%%%%%%%%%%%%%%%%%%%%%%%%%%%%%%%%%% 
\begin{abstract}
%% Word count: 246
The term `history effect' refers to the contribution of any past mass transfer events between a gas bubble and its liquid surroundings towards the current diffusion-driven growth or dissolution dynamics of that same bubble.
The history effect arises from the (non-instantaneous) development of the dissolved gas concentration boundary layer in the liquid in response to changes in the concentration at the bubble interface caused, for instance, by variations of the ambient pressure in time.
Essentially, the history effect amounts to the acknowledgement that at any given time the mass flux across the bubble is conditioned by the preceding time-history of the concentration at the bubble boundary.
Considering the canonical problem of an isolated spherical bubble at rest, we show that the contribution of the history effect in the current interfacial concentration gradient is fully contained within a memory integral of the interface concentration.
Retaining this integral term, we formulate a governing differential equation for the bubble dynamics, analogous to the well-known Epstein-Plesset solution. Our equation 
does not make use of the quasi-static radius approximation. 
An analytical solution is presented for the case of multiple step-like jumps in pressure. The nature and relevance of the history effect is then assessed through illustrative examples.
Finally, we investigate the role of the history effect in rectified diffusion for a bubble that pulsates under harmonic pressure forcing in the non-inertial, isothermal regime.
\end{abstract}

%%%%%%%%%%%%%%%%%%%%%%%%%%%%%%%%%%%%%%%%%%%%%%
%% SECTION: Introduction
%%%%%%%%%%%%%%%%%%%%%%%%%%%%%%%%%%%%%%%%%%%%%%
\section{Introduction} 
\label{sec:introduction}

The diffusion-driven growth and dissolution dynamics of bubbles of a soluble gas are topics that, despite having been studied for a long time, still awaken the interest of scientists and engineers.
In addition to numerous numerical studies on different aspects of the matter, exhaustive analytical treatment in the canonical scenario of the mass (and heat) diffusion-driven growth or dissolution of an isolated bubble has been given over the years. The methods employed are based on the quasi-static approximation \citep{Epstein&Plesset:1950}, thin boundary layer approximation \citep{Plesset&Zwick:1954}, perturbation techniques \citep{Duda&Vrentas:1969}, infinite series \citep{Tao:1978}, integral methods \citep{Rosner&Epstein:1972}, and self-similar solutions for bubble growth starting from zero initial size \citep{Birkhoff:1958, Scriven:1959}, to cite a few.

All these solutions have in common that they assume 
\begin{enumerate}
\item constant ambient pressure during the entire process, and
\item a uniform concentration (or temperature) field in the liquid as initial conditions.
\end{enumerate}
Perhaps the most widely used ones are the Epstein-Plesset solution, valid for growth and dissolution, based on the quasi-stationary approximation \citep{Epstein&Plesset:1950}, and Scriven's exact solution for growth that accounts for the advection term in the diffusion equation \citep{Scriven:1959}. The predicted growth rates have been experimentally validated in supersaturated CO$_2$-water solutions by the works of \cite{Barker:2002} and \cite{Enriquez:2014}, among others. Likewise, in the case of dissolution, experimental verification of the Epstein-Plesset equation has been shown for monocomponent \citep{Kapodistrias&Dahl:2012} and multicomponent \citep{Shim:2014} bubbles.

With respect to the assumption (i) stated above, it happens that the effect of a non-constant pressure-time history has been somewhat overlooked in the derivation of these analytical solutions. In some situations of practical interest, soluble bubbles are subject to successive slow compression-expansion cycles. \cite{Tisato:2015} have successfully proven that the growth-dissolution dynamics of bubbles of a soluble gas can significantly damp the amplitude of seismic waves. 
A second example is the observation of gas bubble disease in stranded cetaceans after being exposed to low-frequency, high-intensity acoustic pulses emitted by sonars. \cite{Crum&Mao:1996} attribute this to bubble growth triggered by rectified diffusion. \cite{Houser:2001} suggested that the likelihood of successful triggering is strongly dependent on the previous history of dives undergone by the cetacean. More specifically, the dive history directly determines the initial supersaturation level of dissolved nitrogen gas in the body fluid surrounding the trapped microbubbles at the time of insonation. 
%{\color{red}
%DELETE: It should be pointed out that the time variations of the pressure considered here occur over timescales much longer than those characteristic of rectified diffusion, where the inertia of the liquid surrounding the bubble must be taken into account. \\(Rectified diffusion does not imply important inertial effects).}

Other scenarios may entail fast, isolated changes in pressure rather than cyclic acoustic forcing. For instance, the growth of gas bubbles trapped in extravascular tissue is aided by steady decompression during depressurisation events \citep{Marzella&Yin:1994}.
\cite{Payvar:1987} studied the mass-transfer-driven bubble growth during rapid liquid decompressions typically found in hydraulic power recovery turbines. A numerical treatment based on the integral method developed by \cite{Rosner&Epstein:1972} was employed. This method requires the imposition of a parabolic concentration profile within a thermal boundary layer of time-varying thickness $L(t)$. \cite{Theofanous:1969} also performed a similar treatment for the growth of vapour bubbles in a superheated fluid under the effect of a decreasing pressure field. 

However, as \cite{Jones&Zuber:1978} pointed out, the integral method is only suitable for a monotonically decreasing pressure-time history and this approach cannot be of general utility. \cite{Jones&Zuber:1978} also studied the growth of vapour bubbles in a superheated liquid under variable pressure. They used the thin boundary layer approximation, which assumes that the temperature spatial variation, $\Delta T$, from the bubble surface temperature to that of the bulk fluid occurs within a thin thermal boundary layer of thickness $L$, much smaller than the bubble radius $R$. As a consequence, the moving bubble boundary could be modelled as a fixed Cartesian plane. The spherical geometry of the bubble was later included in the solution via a correction factor. Little attention was given to the history integral, consequence of the pressure history in time, that appears in their derivations. Instead they focused on bubble growth under a linearly decreasing pressure field. 

While suitable for heat-induced growth, however, in mass-diffusion-driven growth (and specially dissolution) the thin boundary layer approximation \linebreak 
$L(t)/R(t) \ll 1$, may not be always valid. The concentration boundary layer thickness often becomes comparable to the bubble in size, regardless of the value of the diffusion coefficient. To prove this, we can take the well-known asymptotic solutions of \cite{Plesset&Zwick:1954}, \cite{Birkhoff:1958} or \cite{Scriven:1959} for thermal diffusion growth---under assumptions (i) and (ii)---driven by a temperature difference $\Delta T$ between the bubble boundary and the bulk fluid. The bubble radius scales as
\refstepcounter{equation}
$$
R(t) \sim \Ja_\mathit{th} \sqrt{D_\mathit{th}t}, \quad \mbox{with} 
\quad \Ja_\mathit{th} = \frac{\rho_l c_l \Delta T}{\rho_g h_\mathit{fg}}.
\eqno{(\theequation{\mathit{a},\mathit{b}})} 
$$
$D_\mathit{th}$ refers to the thermal diffusion coefficient of the liquid; $\Ja_\mathit{th}$ is the Jakob number for heat transfer, where $c_l$ is the specific heat of the liquid, $\rho_l$ is the liquid density, $\rho_g$ is the gas (vapour) density and $h_\mathit{fg}$ is the latent heat. The thermal boundary layer evolves as $L \sim \sqrt{D_\mathit{th} t}$. It follows that for moderate and high superheats, $L/R \sim \Ja_\mathit{th}^{-1} \ll 1$. 

Equivalently, for mass-diffusion-controlled growth driven by a (molar) concentration difference $\Delta C$ between the bubble boundary and the bulk fluid, \cite{Epstein&Plesset:1950} and \cite{Scriven:1959} among others obtained
\refstepcounter{equation} \label{eq:Jam}
$$
R(t) \sim \Ja_{m} \sqrt{D_{m}t}, \quad \mbox{with} \quad \Ja_{m} = \frac{M_g\Delta C}{\rho_g}.
\eqno{(\theequation{\mathit{a},\mathit{b}})}
$$
$D_m$ denotes to the mass diffusion coefficient, $M_g$ is the gas molar mass and $\Ja_m$ \citep{Szekely&Martins:1971} may be regarded as the analogous Jakob number for mass transfer. For small to moderate supersaturations, at long times the boundary layer thickness is of the order of the bubble radius, $L \sim R$, and analytical treatment accounting for a non-thin boundary layer is therefore essential.
It should be pointed out that the scale laws just described correspond to cases of pure phase-change or mass-transfer driven dynamics. The case where both phenomena are present is more involved and lies beyond the scope of the present work. The interested reader is referred to a recent paper by \cite{Fuster&Montel:2015}.

Turning now to assumption (ii), the effect of the previous growth-dissolution history of the bubble has not, to the best of our knowledge, been explored in detail. 
Naturally, if the bubble has been exchanging mass with its surroundings as a result of previous variations of the ambient pressure, then the concentration field at the beginning of the growth or dissolution stage of interest will not, in general, be uniform. Instead, it is determined by the boundary layer that has grown during that history. The temporal bubble dynamics have been found to be very sensitive to changes in the initial gas concentration profile dissolved in the liquid \citep{Webb:2010}.

\begin{figure}
  \centerline{\includegraphics[width=\textwidth]{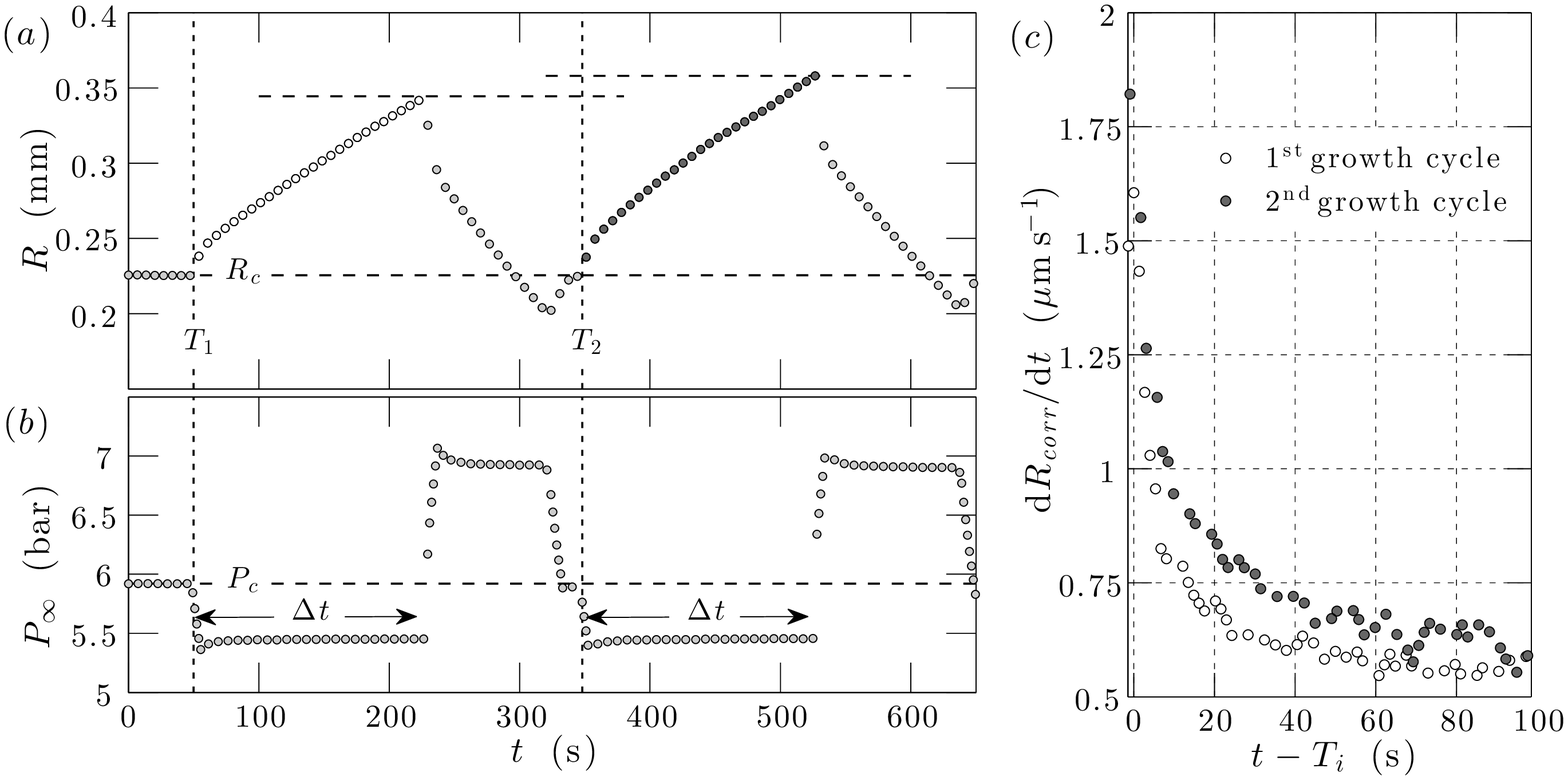}}
  \caption{Experimental plot depicting the growth and dissolution dynamics of a sessile CO$_2$ spherical bubble growing from a 50 $\mu$m pit on a flat chip immersed in a pressurized CO$_2$-water solution. The details of the experiments, performed in the facility described by \cite{Enriquez:2013} and \cite{Enriquez:2014}, will be described in a companion paper. It shows (\textit{a}) the evolution in time of the measured bubble radius $R$ corresponding to (\textit{b}) an imposed variation in the ambient pressure $P_\infty(t)$. The reference pressure, $P_c \!\simeq\! 5.9$ bar, is chosen to be the saturation pressure. Initially, the bubble has a stable radius $R_c \!=\! 225\ \mu$m. At time $T_1$, the ambient pressure is set to drop from $P_c$ to $\simeq \! 5.45$ bar for a period of $\Delta t \!=\! 180$ s. This results in imminent bubble growth (white markers). The same scenario is then repeated at time $T_2$, resulting in a second growth cycle (dark markers). A 5-point moving average filter was performed on the time derivative of the measured radii. This allows a cleaner comparison of (\textit{c}) the rate of growth of the pressure-corrected radius for the two growth cycles. The time axis is initialized on $T_1$ or $T_2$ accordingly. The uncertainty in the growth rate is estimated in $\pm$ 0.0625 $\umu$m/s, much smaller than the differences between both cycles observed initially, at times $t-T_i \lesssim 40$ s.}
\label{fig:intro}
\end{figure}

The effect of the previous history of the bubble on its current dynamics, hereon referred to as the history effect, can be observed from the experimental results in figure \ref{fig:intro}. Figure \ref{fig:intro}(\textit{a}) depicts the evolution of the radius of a bubble under two expansion-compression cycles. Before the first growth cycle at $t = T_1$, the bubble is in equilibrium with its surroundings, implying a uniform concentration field at $t \le T_1$. This is not the case for the second growth cycle. A non-uniform concentration profile is expected at $t=T_2$. Consequently, the measured growth rate is different. In this case it is larger, as will be explained in this paper, hence the bubble grows to a bigger size in the same growth time period.  
The growth rates are shown in figure \ref{fig:intro}(\textit{c}). We have chosen to plot the pressure-corrected radius, defined as
\begin{equation} \label{eq:Rcorr}
R_\mathit{corr}(t) = R(t) \left(\frac{P_\infty(t)}{P_c}\right)^{1/3}.
\end{equation}
This way, the volumetric expansion of the bubble that is solely caused by the pressure drop (a fully non-diffusive effect due to Boyle's law) is removed. The growth rates are initially different, but converge in time. It will be shown in this work that the observed differences in growth rate can be elegantly explained via a history integral of the bubble interfacial concentration, or alternatively, of the ambient pressure.
 
The objective of the present study is twofold:
\begin{enumerate}
    \item to present a theory for the diffusive-driven growth and dissolution of a spherical, isolated bubble that accounts for variable pressure-time history. The dynamics evolve around an expression of similar form to Equation (34) in \cite{Epstein&Plesset:1950} with an additional memory integral term that accounts for the history effect;
    \item to illustrate the importance and nature of the history effect in a couple of bubble mass transfer processes of practical interest through analytical and numerical solutions. 
\end{enumerate}

Attending to these ideas, the paper is structured as follows. In section \ref{sec:histerm}, the mathematical formulation is presented and the history integral term is derived. The governing equation for the bubble dynamics is then developed in section \ref{sec:EPH}. Section \ref{sec:steps} focuses on a bubble exposed to a train of piece-wise constant pressure steps. An analytical solution is presented, in addition to an illustrative example and a brief comparison of the aforementioned solution with numerical simulation.
Section 5 then describes the role of the history effect on the potential growth of an isothermally oscillating bubble under harmonic pressure forcing.
Finally, section \ref{sec:conclusions} summarizes the main conclusions.

%%%%%%%%%%%%%%%%%%%%%%%%%%%%%%%%%%%%%%%%%%%%%%
%% SECTION: Formulation and derivation of the history integral term
%%%%%%%%%%%%%%%%%%%%%%%%%%%%%%%%%%%%%%%%%%%%%%
\section{Formulation and derivation of the history integral term} 
\label{sec:histerm}
%% SUBSECTION: Formulation
\subsection{Formulation}
\label{sec:formulation}

The growth rate of an isolated spherical gas bubble of radius $R(t)$ suspended in a quiescent, infinite liquid environment subject to a time-varying liquid ambient pressure, $P_\infty(t)$, is to be determined. The analysis shall be restricted to monocomponent gas bubbles. The problem has spherical symmetry and consequently $r$ may be taken as the radial distance from the bubble centre, while $t$ is the time variable. 

The molar concentration field $C(r,t)$ of dissolved gas in the liquid is to be solved for from the advection-diffusion equation with spherical symmetry,
\begin{equation} \label{eq:AD}
    \frac{\partial C}{\partial t} + \frac{\dot R  R^2}{r^2}\frac{\partial C}{\partial r} 
    = D_m \frac{1}{r^2} \frac{\partial}{\partial r}\left(r^2 \frac{\partial C}{\partial r}\right),
\end{equation}
where the dot notation stands for $\dd/\dd t$ and $D_m$ is the coefficient of mass diffusion. Notice that the factor $u_r(r,t) = \dot{R}R^2/r^2$ corresponds to the radial velocity field in the liquid that, by virtue of the continuity equation, is induced by the rate of change of the bubble radius.
The concentration field is subject to boundary and initial conditions
\refstepcounter{equation}
$$
    C(R,t) = C_s(t),\qquad
    C(\infty, t) = C_\infty, \qquad
    C(r>R_0, 0) = C_\infty, \qquad   
    \eqno{(\theequation{\mathit{a-}\mathit{c}})}
    \label{eq:bcs_C}
$$
where $R_0 = R(0)$ is the initial bubble radius. The initial concentration of dissolved gas is assumed uniform throughout the liquid and equal to the concentration at the far-field, $C_\infty$.
The concentration boundary condition at the bubble surface is given by Henry's law,
\begin{equation}
    C_s = k_H P_g, 
\end{equation}
where $k_H$ is Henry's coefficient (molar based) and $P_g(t)$ is the total gas pressure inside the bubble. The saturation pressure may thus be defined as $P_\mathit{sat} = C_\infty / k_H$. 
The bubble pressure is determined considering the liquid-gas surface tension $\gamma_{lg}$, but otherwise neglecting the liquid vapour pressure together with damping and inertial effects. 
Naturally, we will not consider the case of an inertially pulsating bubble, driven by strong periodic acoustic forcing \citep{Louisnard&Gomez:2003} or by resonance-triggering frequencies often employed in rectified diffusion.
Indeed, in most diffusion-driven processes, excluding the aforementioned scenario, the contributions of the inertial and viscous terms in the Rayleigh-Plesset equation are negligible since bubble radius accelerations and velocities are relatively small.  
\cite{Payvar:1987} reported that for bubbles of 0.1--1 mm in diameter, $P_\infty(t)$ still remains the dominant term even for fast liquid decompressions of up to 100 bar taking place in less than 1 s.
The bubble pressure is then simply obtained from the Young-Laplace equation,
\begin{equation} \label{eq:yle}
    P_g = P_\infty + \frac{2\gamma_{lg}}{R}.
\end{equation}
An isothermal liquid at temperature $T_\infty$, well below the boiling point, is considered. The bubble volume and pressure are related through the equation of state for an ideal gas,
\begin{equation} \label{eq:igl}
    \frac{4}{3}\upi R^3 P_g = n R_u T_\infty,
\end{equation}
where $n(t)$ is the number of moles inside the bubble and $R_u$ denotes the universal gas constant.
Finally, Fick's first law sets the molar flow rate of gas across the bubble surface to be
\begin{equation} \label{eq:ficks}
    \dot{n} = 4 \upi R^2 D_m \left.\frac{\partial C}{\partial r}\right|_{r = R}.  
\end{equation}
The governing equations (\ref{eq:AD})--(\ref{eq:ficks}) are best treated in dimensionless form. Consequently, let us define the dimensionless radius, ambient pressure, concentration and interfacial concentration as follows:
\refstepcounter{equation}
$$
    a = \frac{R}{R_c}, \qquad 
    p = \frac{P_\infty}{P_c}, \qquad
    c = \frac{C-C_\infty}{k_H P_c}, \qquad 
    c_s = \frac{C_s-C_\infty}{k_H P_c}.
    \eqno{(\theequation{\mathit{a-}\mathit{d}})} \label{eq:nd}
$$
Here, $R_c$ denotes some characteristic bubble radius. Pressure $P_c$ is a characteristic liquid pressure, usually taken as the initial ambient pressure $P_\infty(0)$. 
Additionally, let us introduce the following dimensionless parameters, which remain constant throughout the process:
\refstepcounter{equation}
$$
    \Upsilon = \frac{C_\infty}{k_H P_c}, \qquad
    \Lambda = k_H R_u T_\infty , \qquad
    \sigma = \frac{2\gamma_\mathit{lg}}{R_cP_c}.
    \eqno{(\theequation{\mathit{a-}\mathit{c}})} \label{eq:params}
$$
The parameter $\Upsilon$ refers to the level of saturation of gas in the liquid at the characteristic pressure $P_c$. In fact, the (dimensionless) saturation pressure is simply $p_\mathit{sat} = P_\mathit{sat} / P_c = \Upsilon$. Saturation conditions at $P_c$ are described by $\Upsilon = 1$,  while $\Upsilon <1$ and $\Upsilon > 1$ imply undersaturation and supersaturation respectively. Similarly, note that the saturation level at $P_\infty$ is given by $\Upsilon/p$. Moreover, $\sigma$ represents the characteristic ratio of the Laplace pressure and $P_c$, while $\Lambda$ serves as the solubility parameter. The latter represents the ratio between the bubble's volume and the volume of liquid needed to dissolve, under saturation conditions, the gas it contains.
Henry's law yields a new expression for $c_s$, which, when written in our notation, reads
\begin{equation} \label{eq:c_s}
    c_s = \left(p + \sigma/a\right)-\Upsilon.
\end{equation}
Finally, we shall introduce two different dimensionless time variables, $\tau$ and $\ttau$, in addition to the radial coordinate $\xi$,
\refstepcounter{equation}
$$
    \tau = \frac{D_m}{R_c^2} t, \qquad
    \dd\ttau = \frac{D_m}{R^2(t)}\dd t, \qquad
    \xi = \frac{r}{R(t)}.
    \eqno{(\theequation{\mathit{a-}\mathit{c}})}
$$
The identity 
\begin{equation} \label{eq:tau_id}
   \dd\tau/\dd\ttau = a^2
\end{equation}
directly relates the `physical' dimensionless time $\tau$ and nonlinear time $\ttau$. 
In the upcoming section \ref{sec:history_term}, it will be shown that the mathematical nature of the problem encourages to dispose of the `physical' time $\tau$ and work with the nonlinear time $\ttau$ instead. We advance that this is done in order to analytically solve for the concentration gradient at the bubble boundary from the diffusion equation, all the while accounting for large variations in the bubble radius. Finally, coordinate $\xi \in [1, \infty)$ scales with the instantaneous bubble radius $R(t)$. Hence, the moving bubble boundary is advantageously always mapped by $\xi = 1$. 
Coordinate $\xi$ is an alternative to the Lagrangian coordinate $\eta = \frac{1}{3}(r^3-R^3(t))$ which also eliminates the moving boundary problem. The latter transformation is usually employed in the treatment of rectified diffusion  \citep{Eller&Flynn:1965, Fyrillas&Szeri:1994}. However, in the diffusion-dominant regime of interest, the resulting advection-diffusion equation in $\eta$ is much harder, if not impossible, to treat analytically than the analogous equation in $\xi$.

%% SUBSECTION: The history term on the concentration gradient at the bubble surface
\subsection{The history term on the concentration gradient at the bubble surface}
\label{sec:history_term}
The advection-diffusion equation (\ref{eq:AD}) in dimensionless form becomes
\begin{equation} \label{eq:ADdim}
\frac{\partial c}{\partial \ttau} 
+ \frac{1}{a}\frac{\dd a}{\dd \ttau}  \left(\frac{1}{\xi^2}-\xi \right) \frac{\partial c}{\partial \xi} 
=  \frac{1}{\xi^2} \frac{\partial}{\partial \xi} \left(\xi^2
\frac{\partial c}{\partial \xi} \right).
\end{equation}
Note that the nonlinear time $\ttau$ is the time variable of choice for reasons that will soon become apparent. The boundary and initial conditions (\ref{eq:bcs_C}) of the concentration field $c(\xi, \ttau)$ become
\refstepcounter{equation}
$$
    c(1,\ttau) = c_s(\ttau) = p +\sigma/a -\Upsilon,\qquad
    c(\infty, \ttau) = 0, \qquad
    c(\xi>1, 0) = 0.    
    \eqno{(\theequation{\mathit{a-}\mathit{c}})}
    \label{eq:boundary_conditions}
$$
Only the dimensionless advection term in (\ref{eq:ADdim}) is explicitly dependent on the bubble dynamics through the prefactor $(\dd a/\dd \ttau)/a$. We may characterize this prefactor as a time-dependent P\'eclet number based on the velocity of the bubble boundary,
\begin{equation}
    \Pen(\ttau) = \frac{1}{a}\frac{\dd a}{\dd \ttau}\equiv R\dot R /D_m
\end{equation}
whose magnitude is essentially associated to the instantaneous ratio of advective to diffusive transport.
To illustrate the role of the history effect on the diffusive growth or dissolution rates of bubbles, we will restrict the analysis to mass transfer processes where the advection term in (\ref{eq:ADdim}) is small compared to the diffusion term. We anticipate that the history effect is present in every transient diffusion-advection problem in which boundary conditions change in time. However, should the advective mass transport of species be dominant, the advective nature of the flow would then, at least partially, obscure the contribution of the history effect on the bubble dynamics as well as of any other effect of diffusive nature.

When can we neglect the advection term? 
To find out, we can perform an order of magnitude analysis on (2.12), taking $O(\partial c) = \Delta c$, $O(\xi) = 1$ and denoting $O(\partial\xi)$ as $l$, a characteristic thickness of the boundary layer, defined as $\partial c /\partial\xi|_{\xi=1}  =  \Delta c/l$. The analysis reveals that the diffusion term magnitude is $\Delta c/l^2$, while the advection term has magnitude $O(\Pen)\: \Delta c$.  It follows from the Epstein-Plesset solution that the penetration length of the mass boundary layer will never be much greater than the size of the bubble: $l \lesssim 1$. Thus, the dimensionless advection term may be neglected provided $|\Pen(\ttau)| \ll 1$ at all times. 

The negligible advection assumption is usually valid for small supersaturation or undersaturation ratios, defined as $C_\infty/[k_H P_\infty(t)]-1 \equiv \Upsilon/p -1$. Larger ratios are allowed if the gas solubility is very poor ($\Lambda \ll 1$). In fact, it will be shown in a companion paper that, in the case of a growing or dissolving bubble at pressure $P_\infty$, the P\'eclet number is related to the Jakob number, defined in (\ref{eq:Jam}b), through: $|\Pen| \approx Ja_m^2 = \Lambda|\Upsilon/p -1|$. Thus, a highly soluble gas like CO$_2$ gas in water, ($\Lambda \approx 0.8$ at room temperature) will require that $|\Upsilon/p-1| \ll 1$ for advection to be negligible.

 It must be emphasized that the advection term has two components. The component containing $1/\xi^2$ is the (dimensionless) physical advection term, associated to the radial component of the velocity field,  $u_r(r,t) = R^2\dot R/r^2$. The component containing $-\xi$ is the artificially induced advection that compensates for the scaling nature of $\xi$ with $R(t)$. It naturally arises from the nondimensionalization process. 
Neglecting both terms essentially means that we are now rigorously solving the diffusion equation with an ever-present numerical advection term (associated to the non-physical velocity field $u_r (r,t)= \dot Rr/R$) that accounts for the moving boundary \citep{Penas:2015}. Nonetheless, the effect of the artificial advection on the interfacial concentration gradient will still be small for $|\Pen(\ttau)| \ll 1$.

The concentration field may then be sought by solving  
\begin{equation}
    \frac{\partial c}{\partial \ttau}\label{eq:diff} 
    =  \frac{1}{\xi^2} \frac{\partial}{\partial \xi} \left(\xi^2
\frac{\partial c}{\partial \xi} \right),
\end{equation}
which has no direct dependency on the bubble dynamics.
The choice of $\ttau$ over $\tau$ is justified in that $\ttau$ allows us to arrive from (\ref{eq:AD}) to (\ref{eq:diff}) based on the single assumption that the advection term is small.
Had we chosen to perform an equivalent nondimensionalization with physical time $\tau$ (cf. later Equation (\ref{eq:ad_dim})), it is only possible to treat the arising $a(\tau)$-dependent diffusive term  employing an additional approximation, namely treating the radius $a$ as a constant. 
The essence of the so-called quasi-stationary approximation \citep{Weinberg&Subramanian:1980} behind the Epstein-Plesset equation resides in these two approximations, namely
(1) dropping the advective transport term arising from the interface motion and (2) treating the bubble radius as a constant in the concentration boundary condition at the interface.
Approximation (2), hereon referred to as the {\em quasi-static radius approximation}, is only suitable when considering small or slow changes in the radius size from the equilibrium or initial size. Otherwise, this approximation will inherently decrease the accuracy of the solution. 
It is then concluded that the purpose of $\ttau$ is to avert making use of the quasi-static radius approximation, thereby extending the parameter range in which the theory is valid.
Additionally, writing the advection-diffusion equation in terms of $\tilde{\tau}$ allows us to point out an interesting conclusion regarding the appropriateness of neglecting advection effects. Indeed, the advective term in Equation (\ref{eq:ADdim}) is exactly zero at the bubble interface, as $\left(1/\xi^2 - \xi\right)=0$ there. Thus, in a region close to the bubble surface, it is reasonable to expect the advective term to play a small role even for moderate values of the P\'eclet number.

The solution to (\ref{eq:diff}) for the concentration gradient across the bubble interface (see Appendix \ref{app:historyterm}) reads 
\begin{equation} \label{eq:cgrad}
    -\left.\frac{\partial c}{\partial \xi}\right|_{\xi = 1} = c_s
     +\frac{c_{s0}}{\sqrt{\upi\ttau}} 
     + \int_{0}^{\ttau} \frac{1}{\sqrtsign{\upi(\ttau- \tx)}}
     \frac{\dd c_s}{\dd \tx} \:\dd \tx.
\end{equation}
The initial interfacial concentration, $c_{s0}$, is given by
\begin{equation} \label{eq:c_s0}
    c_{s0} = c_s(0) = p_0+ \sigma/a_0 -\Upsilon,
\end{equation}
where $p_0 = p(0)$ and $a_0 = a(0)$ denote the initial pressure and radius respectively.
The first term in the RHS of (\ref{eq:cgrad}) corresponds to the the steady-state solution determined by the instantaneous interfacial concentration, $c_s(\ttau)$. The second term is the transient component exclusively associated to the initial conditions.
The third and final term is the history or memory integral term. It depends on the time history of $c_s(\ttau)$ caused by any prior variations in the ambient pressure $p(\ttau)$. It addresses the temporal delay required for the concentration boundary layer to become fully developed (physically due to the finite diffusivity $D_m$) as the boundary condition, in this case $c_s$, changes with time.
In other words, the current concentration profile and therefore the mass flux across the bubble are conditioned by the preceding time-history of the boundary condition.

The history integral may be evaluated analytically for the cases when $c_s$ varies in sudden (step-like) jumps and when $c_s$ varies harmonically in time. These will be addressed later in sections \ref{sec:steps} and \ref{sec:oscillations} respectively.

%%%%%%%%%%%%%%%%%%%%%%%%%%%%%%%%%%%%%%%%%%%%%%
%% SECTION: The Epstein-Plesset equation with history term
%%%%%%%%%%%%%%%%%%%%%%%%%%%%%%%%%%%%%%%%%%%%%%
\section{The Epstein-Plesset equation with history term}
\label{sec:EPH}
The ideal gas equation of state provided in (\ref{eq:igl}) may be combined with Fick's law, Equation (\ref {eq:ficks}), to obtain the following mass conservation equation:
\begin{equation} \label{eq:mcons} 
    4 \upi R^2\dot R \left(P_\infty + \frac{2\gamma_{sl}}{R}\right) 
    + \frac{4 \upi R^3}{3} \left(\dot P_\infty - \frac{2\gamma_{sl}}{R^2}\dot R\right)  
    = 4 \upi R^2 \bar{R}T_\infty D \left.\frac{\partial C}{\partial r}\right|_{r = R}.  
\end{equation}
In dimensionless form, it reads
\begin{equation} \label{eq:mconsdim}
     \frac{\dd a}{\dd \ttau}\left(p + \frac{\sigma}{a} \right) + \frac{1}{3}a \left(\frac{\dd p}{\dd \ttau} - \frac{\sigma}{a^2}\frac{\dd a}{\dd \ttau}\right) = \Lambda a \left.\frac{\partial c}{\partial \xi}\right|_{\xi = 1}. 
\end{equation}
  Inserting the expression for the concentration gradient (\ref{eq:cgrad}) into (\ref{eq:mconsdim}) yields the governing equation for the bubble radius dynamics. This may be regarded as the Epstein-Plesset with history term (EPH) equation, written below in terms of $c_s$:
 \begin{equation} \label{eq:EPH} 
     \frac{\dd a}{\dd \ttau}\left(c_s + \Upsilon \right) + \frac{1}{3}a \frac{\dd c_s}{\dd \ttau} 
     = -\Lambda a \left[c_s +\frac{c_{s0}}{\sqrt{\upi\ttau}} 
     +\int_{0}^{\ttau} \frac{1}{\sqrtsign{\upi(\ttau- \tx)}}
     \frac{\dd c_s}{\dd \tx} \:\dd \tx \right].
\end{equation}
The liquid pressure has been related to the surface concentration through $p = c_s + \Upsilon - \sigma/a$ through the relation previously provided in (\ref{eq:c_s}).
In the absence of surface tension, $\sigma = 0$, the EPH equation (\ref{eq:EPH}) may then be conveniently expressed directly in terms of the ambient pressure $p$, 
\begin{equation} \label{eq:EPH_nost} 
     \frac{1}{a} \frac{\dd a}{\dd\ttau}+ \frac{1}{3p} \frac{\dd p}{\dd\ttau} = -\frac{\Lambda}{p} 
     \left[p -\Upsilon
     +\frac{p_0-\Upsilon}{\sqrt{\upi\ttau}} 
     + \int_{0}^{\ttau} \frac{1}{\sqrtsign{\upi(\ttau- \tx)}}
     \frac{\dd p}{\dd \tx} \:\dd \tx  \right].
\end{equation}
To obtain the radius evolution in the physical time, $a(\tau)$, we must numerically integrate the differential EPH equation in (\ref{eq:EPH}) or (\ref{eq:EPH_nost}) for $a(\ttau)$ in addition to the differential equation (\ref{eq:tau_id}) for $\tau(\ttau)$.

%% SUBSECTION: Recovering the Epstein-Plesset equation from the EPH
\subsection*{Recovering the Epstein-Plesset equation from the EPH}
We now show that the Epstein-Plesset equation may be recovered from the EPH equation. In order to do so we revert back to (\ref{eq:mconsdim}). Since the original Epstein-Plesset equation is formulated using the linear time $\tau = D_m t/R_c^2$, it is convenient for both time derivatives in $\ttau$ to be replaced by derivatives in  $\tau$ instead. Recalling that $\dd\tau = a^2\dd \ttau$, we may recast (\ref{eq:mconsdim}) as
\begin{equation}\label{eq:mcons_phys}
     \frac{\dd a}{\dd \tau}\left(p + \frac{2\sigma}{3a} \right)
     + \frac{1}{3}a \frac{\dd p}{\dd \tau} 
     = \frac{\Lambda}{a}\left.\frac{\partial c}{\partial \xi}\right|_{\xi = 1}. 
\end{equation}
The analytical expression for the interfacial concentration gradient is still a function of $\ttau$ as given by (\ref{eq:cgrad}).
The Epstein-Plesset equation assumes a constant pressure-history, i.e. $p = p_0$ and $\dd p/\dd\tau = 0$. 
Equation (\ref{eq:mcons_phys}) then reduces to
\begin{equation} \label{eq:ep1} 
     \frac{\dd a}{\dd \tau}\left(p_0 + \frac{2\sigma}{3a} \right)
     = \frac{\Lambda}{a}\left.\frac{\partial c}{\partial \xi}\right|_{\xi = 1}. 
\end{equation}
The interfacial concentration gradient deserves special treatment.  The history integral term vanishes, consequence of the constant pressure condition. Moreover, unlike the EPH equations (\ref{eq:EPH}) or (\ref{eq:EPH_nost}), the Epstein-Plesset equation is based on the quasi-static radius approximation \citep{Weinberg&Subramanian:1980}. The concentration gradient is calculated for a fixed bubble boundary of radius $a$. The solution is then coupled with the equation for mass conservation (\ref{eq:mcons_phys}), where $a$ is now time-dependent. This has two implications:
(i) the bubble radius $a$ is treated as constant when solving for the concentration gradient. Consequently, (\ref{eq:tau_id}) simplifies to $\tau = a^2\ttau$. Additionally, (ii) the initial interfacial concentration calculated for a static $a$ is in fact `reused' for all its values. This amounts to setting $a_0 = a$ in (\ref{eq:c_s0}), hence $c_{s0} = c_s = p_0 +\sigma/a - \Upsilon$. Under the quasi-static radius approximation, the concentration gradient in (\ref{eq:cgrad}) then becomes
\begin{equation}
    -\left.\frac{\partial c}{\partial \xi}\right|_{\xi = 1} 
    = \left[c_s +\frac{c_{s0}}{\sqrt{\upi\ttau}} \right]
    \approx \left(p_0 -\Upsilon + \frac{\sigma}{a}  \right) \left[1 +\frac{a}{\sqrt{\upi\tau}} \right].   
\end{equation}
Inserting this expression into (\ref{eq:ep1}), one finally recovers the Epstein-Plesset equation in dimensionless form \citep[Equation (34)]{Epstein&Plesset:1950}:
\begin{equation} \label{eq:ep} 
     \frac{\dd a}{\dd \tau} = -\frac{\Lambda(p_0 - \Upsilon +\sigma/a)}{p_0 + 2\sigma/(3a)} 
     \left[\frac{1}{a} +\frac{1}{\sqrt{\upi\tau}} \right]. 
\end{equation}  
An equivalent equation in $\ttau$, which does not make use of the quasi-static radius approximation, may be shown to be
\begin{equation} \label{eq:ep-equiv} 
     \frac{\dd a}{\dd \ttau} = -\frac{\Lambda a}{p_0 + 2\sigma/(3a)}
     \left[ (p_0-\Upsilon) \left(1 +\frac{1}{\sqrt{\upi\ttau}}\right) 
     +\frac{\sigma}{a}\left(1 +\frac{a/a_0}{\sqrt{\upi\ttau}}\right) \right].    
\end{equation}
In the absence of surface tension, $\sigma = 0$, (\ref{eq:ep-equiv}) becomes a separable differential equation in $a$ and $\ttau$. Its analytical solution, with $a(0) = a_0$, is easily found to be
\begin{equation} \label{eq:epn-sol}
a(\ttau) = a_0
\exp \left\{-\Lambda (p_0-\Upsilon) \left( \ttau + 2\sqrtsign{\ttau/\upi} \right)
\right\}.
\end{equation}

The exact analytical solution of $a(\tau)$ to (\ref{eq:epn-sol}) exists in parametric form, with $a(\ttau)$ and $\tau(\ttau)$. The physical time $\tau$ is related to $\ttau$ through
\begin{eqnarray} 
    \tau &=& \frac{a_0}{2 \Lambda (\Upsilon-p_0)}\left[ 
    1-\mathrm{e}^{-2 \Lambda(\Upsilon-p_0)(\ttau + 2\sqrt{\ttau/\upi})}  
    + \sqrt{2\Lambda(\Upsilon-p_0)}\: \mathrm{e}^{2\Lambda(\Upsilon-p_0)} 
    \left\{\mathrm{erf}\left( \sqrtsign{2\Lambda(\Upsilon-p_0)/\upi} \right) \right. \right. \nonumber \\
    &&- \left. \left. \mathrm{erf}\left(\sqrtsign{2 \Lambda(\Upsilon-p_0)\ttau} + \sqrtsign{2\Lambda(\Upsilon-p_0)/\upi} \right) \right\} \right],
\end{eqnarray}
which can be derived by solving (\ref{eq:tau_id}).

%%%%%%%%%%%%%%%%%%%%%%%%%%%%%%%%%%%%%%%%%%%%%%
%% SECTION: Multiple step-like variations in ambient pressure
%%%%%%%%%%%%%%%%%%%%%%%%%%%%%%%%%%%%%%%%%%%%%%
\section{Multiple step-like variations of the ambient pressure}
\label{sec:steps}
This section intends to shed light on the history effect in bubble growth through illustrative examples based on analytical and numerical solutions. To this end, we shall consider the case where the time-dependent ambient pressure consists of $N$ consecutive step-like jumps in pressure. At the $n$-th jump taking place at nonlinear time $\tT_n$, the pressure changes by $\Delta p_n$. The pressure history and its time derivative may be modelled as
\refstepcounter{equation}
$$
    p = p_0+ \sum_{n = 1}^N \Delta p_n H(\ttau- \tT_n), \qquad
    \frac{\dd p}{\dd \ttau} = \sum_{n = 1}^N \Delta p_n \delta(\ttau- \tT_n), \quad \mbox{for} \quad n = 1,\ldots N,
    \eqno{(\theequation{\mathit{a},\mathit{b}})} \label{eq:pstep}
$$
where $H$  denotes the Heaviside function and $\delta$ is the Dirac delta.
As anticipated at the end of section \ref{sec:histerm}, it is then possible to analytically evaluate the history integral term and consequently solve the EPH equation provided the Laplace pressure is neglected. 
The analytical solutions to the EPH Equation (\ref{eq:EPH_nost}) and Equation (\ref{eq:diff}) for the concentration field are derived in Appendix \ref{app:solution_steps}.

\subsection{An example to illustrate the history effect} 
\label{sec:example}
\begin{figure}
  \centerline{\includegraphics[width=\textwidth]{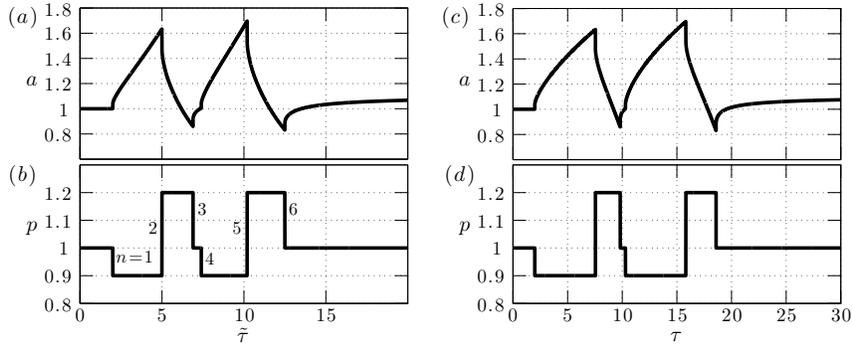}}
  \caption{Analytical solution for (\textit{a})  the evolution of the dimensionless bubble radius, $a$, corresponding to (\textit{b}) multiple jumps in the dimensionless ambient pressure $p$, both plotted against the dimensionless, nonlinear time $\ttau$. Initial conditions correspond to perfect saturation.
Additionally, equivalent plots of (\textit{c}) the dimensionless bubble radius and (\textit{d}) dimensionless ambient pressure are plotted against the dimensionless linear time $\tau$. For reference, the physical parameters employed are those considering a CO$_2$ bubble of size $R_c \!=\! 225 \ \mathrm{\mu m}$ in water 
($k_H \!=\! 3.40$$\times$$10^{-4}\  \mathrm{mol\: N^{-1}\: m^{-1}}$, 
$D_m \!=\! 1.92$$\times$$10^{-9} \ \mathrm{m^2\: s^{-1}}$) under conditions 
$P_c \!=\! 4.9$ bar and $T_\infty \!=\! 293$ K. Surface tension is neglected. The resulting solubility and saturation parameters are $\Lambda \!=\! 0.828$ and $\Upsilon \!=\! 1$.}
\label{fig:example1}
\end{figure}

Let us consider the radius history of a bubble with negligible Laplace pressure ($\sigma = 0$) initially in equilibrium at the saturation pressure. Setting $P_c = P_\infty(0)$ and $R_c = R(0)$ renders $p_0 = 1$, $a_0 = 1$ and $\Upsilon =1$. The bubble is then exposed to $N = 6$ consecutive jumps in pressure. We prescribe coefficients $\tT_n$ and $\Delta p_n$, tabulated in table \ref{tab:pcoeffs}. These are entered into (\ref{eq:pstep}), resulting in the pressure-time history plotted in figure \ref{fig:example1}(\textit{b}). An iterative procedure was employed to establish coefficients $\tT_n$  such that the two expansion (growth) stages ($T^+_1 < \tau < T^-_2$ and $T^+_4 < \tau < T^-_5$) have the same duration in $\tau$ (see figure \ref{fig:example1}\textit{d}). 
Moreover, both expansion stages are identical in the sense that $a(T_1) = a(T_4) = a_0 = 1$, $\Delta p_1 = \Delta p_4$ and $p_1 = p_4$.  The analytical solution for $a(\ttau)$ provided by (\ref{eq:anaEPH1}) and (\ref{eq:anaEPH2}) is plotted in figure \ref{fig:example1}(\textit{b}). 
The solution is more naturally interpreted when presented in $\tau$ (figure \ref{fig:example1}\textit{c}). This requires numerical integration of (\ref{eq:tau_id}) once the analytical solution for $a(\ttau)$ has been obtained. 

At $\tau = T^-_1$, right before the first growth stage, the history integral is identically zero, i.e. there is no previous history. This is obviously not the case at $\tau = T^-_4$. Consequently, the memory effect must be behind the differences in bubble growth observed between these two pressure-wise identical expansion periods (figures \ref{fig:example1}\textit{a,c}). 

\begin{table}
  \begin{center}
\def~{\hphantom{0}}
  \begin{tabular}{c c c c c c c}
       $n$        &1        &2        &3        &4      &5       &6    \\ [3pt]
       $\tT_n$    &2        &5        &6.9      &7.4    &10.2    &12.5   \\ 
       $T_n$      &2        &7.5     &9.8       &10.3   &15.8    &18.6   \\     
    $\Delta p_n$  &-0.1     &0.3      &-0.2     &-0.1   &0.3     &-0.2   \\     
    \end{tabular}
    \caption{Coefficients $\tT_n$ and $\Delta p_n$ of the pressure step function used in the example (see figure \ref{fig:example1}). The resulting coefficients $T_n$ associated to the physical time $\tau$ are also included.}
    \label{tab:pcoeffs}
  \end{center}
\end{table}

\begin{figure}
  \centerline{\includegraphics[width= 0.95\textwidth]{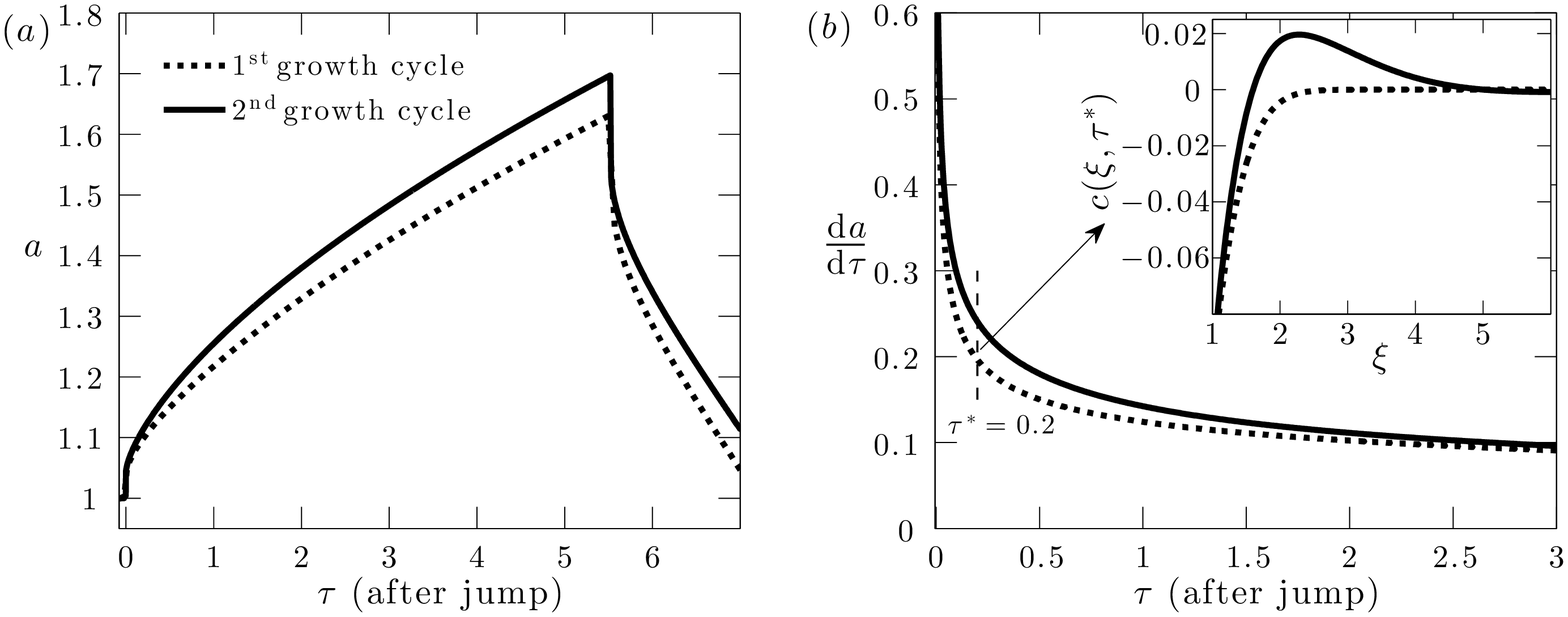}}
  \caption{The first (dashed line) and second (solid line) growth cycles of the bubble (see figure \ref{fig:example1}) are compared through (\textit{a}) the dimensionless radius $a$  and (\textit{b}) its rate of change, computed numerically out of the analytical $a(\tau)$. These are plotted against the dimensionless linear time $\tau$ after each of the two negative pressure jumps ($\Delta p_1$ and  $\Delta p_4$) that lead to undersaturation. Inset: dimensionless concentration radial profiles evaluated at a short time $\tau^*$ immediately after each of the two jumps.}
\label{fig:example_comp}
\end{figure}

To highlight this effect, figure \ref{fig:example_comp}({\it a}) compares the bubble radius history during both growth stages. At equal times, measured from the beginning of each expansion period, the bubble always exhibits a larger radius in the second growth cycle. This is consequence of the larger, history-augmented growth rate (figure \ref{fig:example_comp}$b$). The second growth rate is most enhanced at the beginning. It then asymptotically converges in time with the first growth rate, by then confirming the complete dissipation of the history effect.

The physical explanation of the history effect lies in the concentration profiles near the bubble a short time after each jump (figure \ref{fig:example_comp}{\it b}).
Both concentration profiles are bounded by an identical value of $c_s(\tau^*) <0$, and the far field concentration $c(\infty, \tau) = 0$. 
The first concentration profile corresponds to a uniform initial condition $c(\xi, T^-_1) = 0$. On the other hand, the second concentration profile remembers that $c_s >0$ during the previous dissolution period at $T^+_2 < \tau < T^-_3$.
Consequently there exists a supersaturation region ($c>0$) near the bubble, containing the mass of gas that was transferred to the liquid during said dissolution period. The concentration boundary layer takes time to fully develop in response to any change in the interfacial concentration. This effective adaptation time, owing to the finite diffusivity, results in this case in a momentarily steeper interfacial concentration gradient that enhances the growth rate substantially.

The effect described here is also present in the experimental data shown in figure \ref{fig:intro}. As a matter of fact, both the size and growth rate time-histories are qualitatively identical to those shown in figure \ref{fig:example1}, predicted by the analytical solution. The quantitative analysis of these experiments will be carried out in a future companion paper, as it involves taking into account the presence of the plate and natural convection, among other effects.

Another manifestation of the history effect is observed for $\tau > T^+_6$, after the expansion-compression cycles have ended (see figure \ref{fig:example1}). Even though the pressure returns to the saturation value, $c_s(\tau \geq T^+_6) = 0$, static equilibrium is not instantaneously achieved. Like before, a supersaturation region near the bubble ($c>0$) remains from the preceding dissolution stage. This amounts to a positive interfacial concentration gradient. The radius of the bubble grows from  $a(\tau = T^+_6) \approx 0.9$ to $a(\tau = 30) \approx 1.1$, certainly a non-negligible increment. This growth is entirely provided by the history effect. In other words, the interfacial concentration gradient in (\ref{eq:cgrad}) is non-zero due to the sole contribution of the history integral term. 

%% SUBSECTION: Advection and surface tension effects
\subsection{Comparison with numerical simulation}

An important question that arises now is how important are history effects compared to others that have been neglected in our discussion thus far, such as advection or surface tension. To shed light on this question, we numerically solve the full mass-transfer problem (\ref{eq:ADdim}), (\ref{eq:boundary_conditions}) and (\ref{eq:mconsdim}) which takes these effects into account. 
Details of the numerical treatment of the problem may be found in Appendix \ref{app:numerical_model}.

\begin{figure}
  \centerline{\includegraphics[width= 0.95\textwidth]{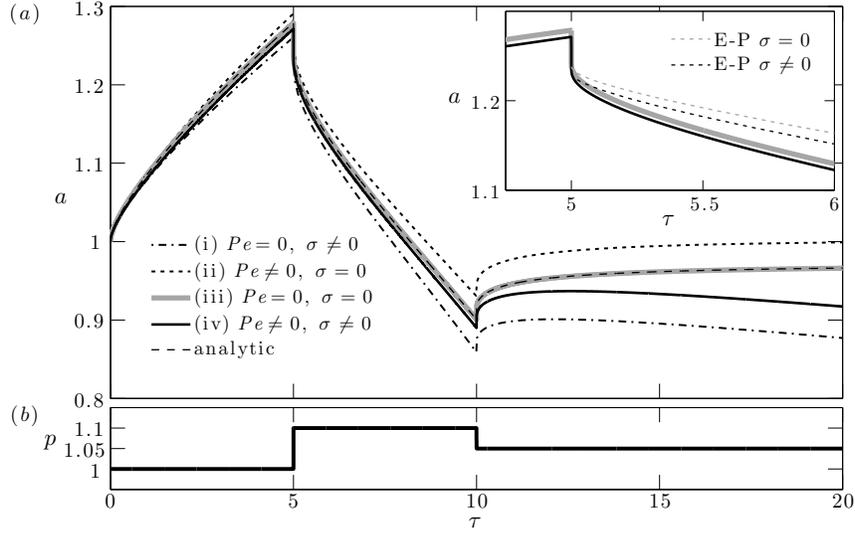}}
  \caption{Evolution of (\textit{a}) the dimensionless bubble radius $a$ in time $\tau$ according to the simulation scenarios (i)--(iv) and the analytical solution in (\ref{eq:anaEPH1}) and (\ref{eq:anaEPH2}). The bubble is initially in a 105\% supersaturated liquid. It is exposed to (\textit{b}) a prescribed pressure-time history consisting of two particular jumps. Inset: the dissolution dynamics for $\tau \!>\!5$ according to the Epstein-Plesset solution (\ref{eq:ep}) are compared with solutions (iii) and (iv).
For reference, the physical parameters employed are those considering a CO$_2$ bubble of size $R_c \!=\! 175 \ \mathrm{\mu m}$ in water 
($k_H \!=\! 3.40$$\times$$10^{-4}\  \mathrm{mol\: N^{-1}\: m^{-1}}$, 
$D_m \!=\! 1.92$$\times$$10^{-9} \ \mathrm{m^2\: s^{-1}}$,
$\gamma_\mathit{lg} \!=\! 0.07 \ \mathrm{N\: m^{-1}}$) under conditions 
$P_c \!=\! 2$ bar and $T_\infty \!=\! 293$ K.
The resulting surface tension, solubility and saturation parameters are $\sigma \!=\! 0.004$, $\Lambda \!=\! 0.828$ and $\Upsilon \!=\! 1.05$.}
\label{fig:example_simulation}
\end{figure}

As an example, let us consider a bubble in an initially 105\% supersaturated liquid. The initial concentration field of dissolved gas is assumed uniform.
Setting $P_c = P_\infty(0)$ and $R_c = R(0)$ entails $p_0 = 1$, $a_0 = 1$ and $\Upsilon = 1.05$. The bubble is then exposed to the pressure-time history shown in figure \ref{fig:example_simulation}({\textit b}). There are three constant-pressure stages, the first corresponding to liquid supersaturation (expansion stage), the second to undersaturation (compression stage) and the third to perfect saturation ($p_\mathit{sat} = \Upsilon = 1.05$). The set of physical conditions are specified in the caption of figure \ref{fig:example_simulation}. Four different versions of the problem are numerically solved, namely
\begin{enumerate}
    \item neglecting the dimensionless advection term, i.e. setting $\Pen = 0$ in (\ref{eq:ADdim2});
    \item neglecting surface tension, i.e. setting $\sigma = 0$ in (\ref{eq:a'sim}) and (\ref{eq:boundary_conditions_numerical});
    \item neglecting both the advection term and surface tension (the same scenario as the one analytically solved, i.e. $\Pen = 0$, $\sigma = 0$) and
    \item taking into account both the advection term and Laplace pressure.
\end{enumerate}
These solutions are compared in figure \ref{fig:example_simulation}({\textit a}), along with the analytical solution (for which $\Pen = \sigma = 0$). 
History effects are critical in order to properly describe this particular dissolution (compression) stage ($5 <\tau <10$). While growth is well determined by existing solutions \citep{Epstein&Plesset:1950, Scriven:1959}, the dissolution experienced by this bubble is greatly affected by the low-concentration boundary layer left by the preceding growth stage. It stands to reason that at the initial instants of dissolution, during which history effects are most important, the dissolution rate observed is much faster than that obtained under the assumption of a historyless, uniform concentration field at $\tau = 5^+$. This is corroborated in the inset plot of figure \ref{fig:example_simulation}, where we plot the 
Epstein-Plesset solution (\ref{eq:ep}), which assumes a uniform initial concentration field (with $p_0 = 1.1$). To enable proper comparison, the initial radius $a_0$ at $\tau = 5^+$ for the Epstein-Plesset solution with and without surface tension has been fitted to that predicted by curve (iv) and (iii) respectively. Note that curve (iii) is identical to the analytical solution. 

When the pressure drops to the saturation value at $\tau = 10$, the bubble continues to grow for some time. As discussed in section \ref{sec:example}, it is consequence of the high concentration boundary layer of dissolved gas in the nearby liquid that was left by the preceding dissolution stage. Since the bubble is nearly close to the equilibrium in this stage, both advection and surface tension are small compared to history, at least for a few time units. At later times though, surface tension breaks the diffusive equilibrium, and ultimately drives the bubble towards its total dissolution. Nonetheless, this occurs at times much longer than those of the pressure cycle considered here.

Under the conditions investigated here, the effect of advection and surface tension on the bubble radius dynamics nearly cancel each other out during the expansion-compression cycle. Consequently, it is observed that the full solution (iv) (which contains both surface tension and advection) is in better agreement with the analytical solution (iii) (which contains neither) than with curves (i) or (iii) during this period. In fact, a quantitative explanation may be obtained from a simple analysis.
For small values of $\Pen$ and constant ambient pressure, the contribution of advection towards the growth or dissolution rate may be quantified using the asymptotic solution of \cite{Duda&Vrentas:1969}, Equation (43). To do so, we subtract the Epstein-Plesset solution (\ref{eq:ep}) with $\sigma = 0$ from Equation (43). For our particular set of initial conditions, $p_0 = a_0 = 1$, we obtain 
\begin{equation}
\left[\frac{\dd a}{\dd\tau}\right]_{\Pen \neq 0} - 
\left[\frac{\dd a}{\dd\tau}\right]_{\Pen = 0}
 \approx \Lambda^2\left(1-\Upsilon\right)^2\left[1-\frac{2}{\upi}
 +\frac{1}{\sqrtsign{\upi\tau}}\left(\tau + \frac{2}{\upi}-\frac{1}{2}\right)\right].
\end{equation}
Conversely, a simple estimation of the effect of the Laplace pressure on the rate of change of the bubble radius can be done by subtracting (\ref{eq:ep}) with $\sigma = 0$ from itself. Provided $p_0 = a_0 = 1$ and $\sigma \ll 1$, this results in  
\begin{equation}
\left[\frac{\dd a}{\dd\tau}\right]_{\sigma \neq 0} - 
\left[\frac{\dd a}{\dd\tau}\right]_{\sigma = 0} \approx
-\frac{\Lambda\sigma}{3a}(1+2\Upsilon)
\left(\frac{1}{a}+\frac{1}{\sqrtsign{\upi\tau}}\right).
\end{equation}
We may use these formulas to evaluate the characteristic porcentual contribution to the growth rate of each effect with respect to that given by Epstein-Plesset solution (\ref{eq:ep}) with $\sigma=0$. Halfway through the initial growth stage ($\tau = 2.5$, $a = 1.18$) we determine that advection enhances growth by about 5\% whereas surface tension slows it down by $-7\%$. Thus, this small net difference of 2\% (which actually decreases as the bubble grows) justifies the close agreement between curves (iii) and (iv).

%%%%%%%%%%%%%%%%%%%%%%%%%%%%%%%%%%%%%%%%%%%%%%
%% SECTION: The history effect in small amplitude oscillations
%%%%%%%%%%%%%%%%%%%%%%%%%%%%%%%%%%%%%%%%%%%%%%
\section{Small amplitude isothermal oscillations}
\label{sec:oscillations}

This section aims to deliver insight on the role of the history effect in the problem concerning the mass transfer across a bubble that pulsates non-inertially under sinusoidal acoustic excitation. This constitutes a particular regime under the broad phenomenom known as rectified diffusion, associated with many practical applications. For a more general description of this complex phenomenon the reader is referred to the seminal works of \cite{Hsieh&Plesset:1961}, \cite{Eller&Flynn:1965}, \cite{Crum&Hansen:1982} or the more recent work by \cite{Zhang&Li:2014}. Here, we will only consider low frequency forcing, which has special relevance in seismological events \citep{Tisato:2015}, and in the exposure of marine mammals to low-frequency sonars \citep{Crum&Mao:1996}.
More specifically, we shall restrict the present analysis to small amplitude, isothermal oscillations in consistency with Equations (\ref{eq:yle}) and (\ref{eq:igl}). 

Provided the bubble pressure varies harmonically in time, it is then possible to readily determine the role of the history effect on the mass transfer across such bubble. In turn, it may be used to explain and predict the nature (phase and amplitude) of oscillation. To this end, let us consider a harmonic pressure forcing,  
\begin{equation} 
    P(t) = P_c \left[ 1 + \varepsilon \sin(2\upi f_c t)\right],
\end{equation} 
where $\varepsilon$ is the dimensionless forcing amplitude and $f_c$ is the forcing frequency.
This time it shall prove convenient to work with the linear time $\tau = D_m t/R_c^2$. The dimensionless pressure $p(\tau)$ is then
\refstepcounter{equation}
$$
    p(\tau) = 1 + \varepsilon \sin\Omega\tau,\qquad \mbox{with} \quad
    \Omega = \frac{2\upi f_c R_c^2}{D_m}.
    \eqno{(\theequation{\mathit{a},\mathit{b}})}
$$
The angular forcing frequency $\Omega$ has been non-dimensionlized with the diffusive timescale. The frequency $\Omega/(2\upi)$ is in fact equivalent to the frequency-based P\'eclet number typically used in rectified diffusion problems, $\Pen_f = f_c R_c^2/D_m$, as introduced by \cite{Fyrillas&Szeri:1994}.
We propose a solution for the bubble dynamics of the form
\begin{equation} \label{eq:a_def}
    a(\tau) = \bar a(\tau) - \delta \sin{(\Omega \tau + \phi)} + O(\delta^2),
\end{equation}
where $\bar a = \bar R / R_c$ is the dimensionless equilibrium radius, while $\delta$ is the oscillation amplitude. Lastly, $\phi = \phi(\Omega)$ denotes the phase shift in oscillations compared to the phase expected for an equivalent non-soluble, isothermally contracting and expanding bubble under the sole effect of the oscillatory ambient pressure forcing. 

Solution (\ref{eq:a_def}) and the analysis soon to follow make use of three underlying assumptions:
\begin{enumerate}
\item[(a)]{the bubble remains isothermal throughout the oscillations. Inertial and viscous effects in the bubble dynamics are completely neglected. The effect of the Laplace pressure on the oscillatory problem is also ignored: $\sigma/\bar a \ll 1$.}
\item[(b)] the strain amplitude of the oscillations is small, $\delta/ \bar a \ll 1$;
\item[(c)] the equilibrium radius $\bar a(\tau)$ is assumed to vary sufficiently slowly in time to be treated as constant within an individual oscillation period. This timescale for bubble growth/dissolution must be much larger than the period of oscillation, $\Omega^{-1}$. 
\end{enumerate}

The derivation and discussion of the range of frequency $\Omega$, pressure amplitude $\varepsilon$ and saturation level $\Upsilon$ for which assumptions (a)--(c) hold is provided in Appendix \ref{app:conditions}.
Here, we choose to highlight that assumption (a) requires that the value of $\Omega$  must satisfy
\refstepcounter{equation}\label{eq:omega_range}
$$
\Omega < D_\mathit{th}/D_m, \qquad 
1 \lesssim \Omega \ll \Omega_\mathit{res}    \eqno{(\theequation{\mathit{a},\mathit{b}})}.
$$
where $D_\mathit{th}$ is the thermal diffusivity of the gas at constant volume and $\Omega_\mathit{res}$ is the dimensionless Minnaert resonance frequency. Note that this range may span several orders of magnitude. Moreover, assumption (b) requires that $ \varepsilon \ll 1$.

Assumption (a) allows for the gas pressure in the bubble to be the same as the ambient pressure $p(\tau)$ and consequently $c_s(\tau) = p(\tau) - \Upsilon$.
The radius $a_p$ of the equivalent non-soluble bubble mentioned above (for which $\phi_p \equiv 0$) then satisfies $p\:a_p^3 = \bar a^3$.  Linearization results in
\begin{equation} \label{eq:ap_def}
a_p(\tau) = \bar a p^{-1/3} = \bar a(1+\varepsilon\sin\Omega\tau)^{-1/3} = \bar a -\frac{\varepsilon \bar a}{3}\sin\Omega\tau + O(\varepsilon^2\bar a^2).
\end{equation}
Naturally, (a) implies that $a_p$ is in anti-phase with pressure $p$ and interfacial concentration $c_s$. 

To conclude, we would like to remark that assumptions (a)--(c) essentially simplify this complex mass transfer problem to the point that the role of the history effect can be smoothly rooted out and analyzed by means of simple analytical expressions. This is the purpose of the next subsection.

We anticipate that it is beyond the scope of this work to determine the threshold pressure amplitude for growth under rectified diffusion or to model the bubble growth rate under specific low-frequency scenarios of interest. These notions have been investigated in the case of volcanic systems \citep{Ichihara&Brodsky:2006}, and in high supersaturation levels with potentially large amplitude oscillations as observed in the capillaries of marine mammals \citep{Crum&Mao:1996, Ilinskii:2008} in which the effect of the viscoelastic medium is undoubtedly important \citep{Zhang&Li:2014b}.

\subsection{The oscillatory problem} 
\label{sec:oscillatory_problem}

The full advection-diffusion problem is usually tackled by splitting it into a smooth problem and an oscillatory problem following the work of \cite{Fyrillas&Szeri:1994}. 
The smooth problem (diffusion timescale $\tau \sim 1$)
accounts for the steady part of the boundary condition and yields the net flux of dissolved gas across the oscillating bubble. The oscillatory problem (diffusion timescale $\tau \sim \Omega^{-1}$) takes into account the unsteady part of the boundary condition and describes the zero-average-mass exchange occurring over one bubble oscillation. This approach is only valid on the assumption that the timescales of the two processes are well separated: $\Omega \gg 1$. Since we are also considering frequencies $\Omega \sim 1$, splitting the problem is no longer possible. 
Moreover, the smooth solution, whether following the approach of \cite{Eller&Flynn:1965} or \cite{Fyrillas&Szeri:1994} is constructed from the mass transport equation in  Lagrangian spherical coordinates (as opposed to (\ref{eq:ad_dim})) and requires special time averaging of the radius dynamics to properly capture the effect of advection. The interested reader is referred to \cite{Ilinskii:2008} for a brief review on these two forms of the smooth solution. 

Fortunately, the history effect is only directly relevant in the oscillatory problem associated to the unsteady boundary condition.
In other words, we are just interested in the solution during an individual period of oscillation about $\bar a$. Solving the governing mass transport equation directly in the form 
\begin{equation} \label{eq:ad_dim}
\frac{\partial c}{\partial \tau} +\frac{\Pen(\tau)}{a^2} \left( \frac{1}{\xi^2}-\xi \right) \frac{\partial c}{\partial \xi}  = \frac{1}{a^2 \xi^2}\frac{\partial}{\partial \xi}\left(\xi^2 \frac{\partial c}{\partial \xi} \right) 
\end{equation} 
is suitable to this end. Note that $\Pen(\tau) = (\dd a/\dd \tau) \: a \equiv R \dot R/D_m$ is our $\dot R$-based P\'eclet number (as previously defined in Section 2.2). 
It follows that under the small amplitude oscillation restriction, the magnitude of the dimensionless advection term in (\ref{eq:ad_dim}) is always smaller than the unsteady and diffusive terms by a factor of $O(\varepsilon)$, where $\varepsilon \ll 1$ (see Appendix \ref{app:oscillatory}).
Note that this condition holds true regardless of the magnitude of $\Pen$ (cf. $O(\Pen) \sim \varepsilon\Omega \bar a^2/3$), which may not necessarily be smaller than unity. It is therefore not unreasonable to neglect the advection term when dealing just with the oscillatory problem consisting of an individual cycle.
Note, nonetheless, that the small contribution of the advection term on the interfacial gradient (negligible over one individual cycle) is still crucial to provide net growth over many cycles.

Moreover, assumption (b) allows to make the approximation $a(\tau) = \bar a(\tau) + O(\delta)  \approx \bar a$, which is constant over an oscillation period according to (c). While this greatly simplifies the problem, it only induces very small errors of $O(\varepsilon) \ll 1$.  

Additionally, we will consider saturation conditions $\Upsilon = 1$. This is justified in that the bulk concentration of the liquid comes into play in the smooth solution only \citep{Fyrillas&Szeri:1994}. Similarly, surface tension and time averaged bubble interface motion are mostly relevant in the smooth solution.

Implementing these notions , the solution to (\ref{eq:ad_dim}) for the concentration profile is found to be (see Appendix \ref{app:oscillatory} for the complete derivation)
\begin{equation} \label{eq:c_profile}
c(\xi,\tau) = \frac{\varepsilon}{\xi}\exp\left\{-\bar a\sqrtsign{\Omega/2}(\xi-1)\right\}
\sin\left\{\Omega\tau - \bar a\sqrtsign{\Omega/2}(\xi-1)\right\}
\end{equation}
which renders the following concentration gradient at the wall:
\begin{equation} \label{eq:cgrad_wall}
    \left.\frac{\partial c}{\partial \xi}\right|_{\xi = 1} 
     = -\varepsilon \left[\sin\Omega\tau + {\bar a}\sqrt\Omega\sin\left(\Omega\tau + \frac{\upi}{4}\right) \right].
 \end{equation}
The first term in the RHS of (\ref{eq:cgrad_wall}) refers to the quasi-steady mass flux imposed by the current value of $c_s(\tau) = \varepsilon \sin\Omega\tau$. 
The last term in the RHS constitutes the contribution from the history integral (derived in Appendix \ref{app:oscillatory}).
In the scenario of small amplitude oscillations considered here, the mass flux coming from the contribution of the history term is thus observed to be proportional to the pressure amplitude $\varepsilon$. The history effect imposes a phase shift in the sinusoidal interfacial concentration gradient, $0< \phi_\mathit{grad}(\Omega) \leq \upi/4$, with respect to the phase of $-c_s(\Omega\tau)$ or the oscillatory component of $a_p(\tau)$.

As $\Omega \rightarrow 0$, the history integral term vanishes and $\phi_\mathit{grad} \rightarrow 0$. The mass flux is in phase with $a_p$ (cf. (\ref{eq:ap_def})).
Conversely, as $\Omega \rightarrow \infty$, the instantaneous mass flux is entirely provided by the history integral. The maximum possible phase shift with respect to $a_p$ is attained: $\phi_\mathit{grad} = \upi/4$.
In fact, the frequency dependency of this phase shift is independent of $\varepsilon$ and may be shown to be
\begin{equation}\label{eq:phi_grad}
\phi_\mathit{grad}(\Omega, \bar a) = -2\arctan\left\{1+\frac{\sqrtsign{2}}{\bar a\sqrtsign\Omega} -  \frac{\sqrtsign{2}}{\bar a\sqrtsign\Omega}
\left( \bar a^2 \Omega + \sqrtsign{2\Omega}\bar a + 1 \right)^{1/2} \right\}.
\end{equation}

\begin{figure}
  \centerline{\includegraphics[width= 0.95\textwidth]{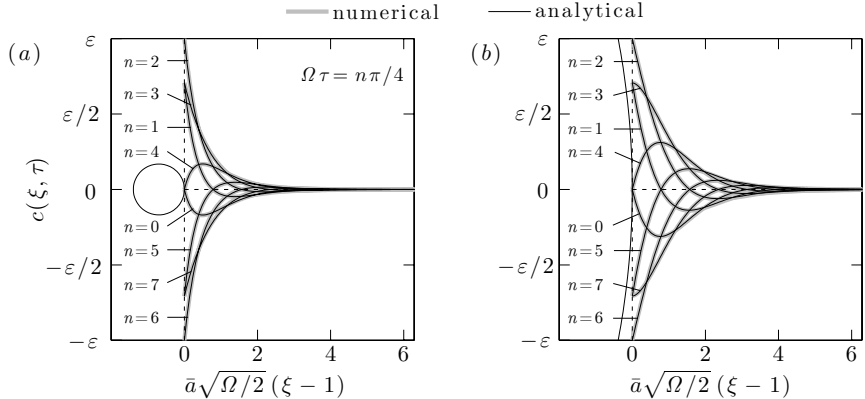}}
  \caption{Radial concentration profiles at different times over an oscillation period for (\textit{a}) $\Omega \!=\! 1$  (\textit{b}) $\Omega \!=\! 1000$ taking $\bar a\! =\! 1$ and $\varepsilon \!=\! 0.01$. The radial coordinate has been normalized by the wavenumber and spans one wavelength. The circle represents the relative bubble size (drawn to scale with the horizontal axis) for reference.}
\label{fig:profiles}
\end{figure}

Solution (\ref{eq:c_profile}) may be described as a damped transverse wave with wavenumber $k = \bar a\sqrtsign{\Omega/2}$ and phase velocity $\sqrtsign{2\Omega}/\bar a$ attenuating over a penetration depth $l \sim 1/(\bar a\sqrt\Omega)$. 
Figure \ref{fig:profiles} shows the close agreement of this (advectionless) analytical solution  with the full numerical computation (taking into account advection) involving (\ref{eq:ad_dim}) and the mass conservation equation (\ref{eq:mcons_phys}) with no surface tension for $\Omega = 1$ and $\Omega = 1000$ (with equilibrium radius $\bar a = 1$). This verifies the negligible qualitative impact of the advection term in the oscillatory problem even for this moderate pressure amplitude of $\varepsilon = 0.01$. On another note, the boundary layer thickness $h$ may be best estimated as the wavelength, $h = 2\upi/k$. Over  a distance $h$, (horizontal span in figure \ref{fig:profiles}) the concentration amplitude reduces by a factor of $\mathrm{e}^{2\upi} \approx 540$. 

For $\Omega = 1000$, figure \ref{fig:profiles}(\textit{b}) hints that the shift in the phase of the concentration gradient with respect to $-c_s$ is practically that of the high frequency limit, $\phi_\mathit{grad} = \upi/4$. The interfacial gradient is flat at $\Omega\tau = 3\upi/4,\: 7\upi/4$ as opposed to $c_s(\tau) = 0$ taking place at $\Omega\tau = \upi, \:2\upi$. For $\Omega =  1$,  direct evaluation of (\ref{eq:phi_grad}) yields $\phi_\mathit{grad} = \upi/8$ (as later observed in figure \ref{fig:waveform_omega1}{\textit a}).

It stands to reason that solutions (\ref{eq:c_profile}) and (\ref{eq:cgrad_wall}) evaluated in the limit $\Omega \gg 1$ converge to those one may obtain following the approach of \cite{Fyrillas&Szeri:1994} (inherently valid for $\Omega \gg 1$) subject to the quasi-static radius approximation, i.e. taking $a(\tau) = \bar a$ as constant. The complete derivation may be found in Appendix \ref{app:fs}.   

\subsection{Strain amplitudes and phase}
The frequency-dependent $\phi_\mathit{grad}$ induced by the history effect implies that pressure-induced contraction of the bubble radius ($p>1$, $c_s>0$, $\dd a_p / \dd\tau<0$) is not in phase with dissolution (negative mass diffusion rate, $\partial c/\partial \xi(1, \tau)>0$).  Likewise, the pressure-induced expansion is not in phase with growth. We would like to conclude this work by determining the dependency on $\Omega$ of the overall strain amplitude and phase lag with respect to $a_p$ $(\phi<0)$ of the isothermally pulsating bubble. 

Let us then define the dimensionless pressure corrected radius, $\acorr(\tau)$,  in an analogous manner to $R_\mathit{corr}$ in (\ref{eq:Rcorr}) as follows:
\begin{equation} \label{eq:acorr_def}
    a = p^{-1/3}\:\acorr =  a_\mathit{p} \:\acorr /\bar a.
\end{equation}
The pressure corrected radius is associated purely with the mass transfer across the interface. It has the useful property that its rate of change, $\dd\acorr/\dd\tau$, always takes the same sign as the interfacial concentration gradient.Consequently, $\phi_\mathit{corr} = \phi_\mathit{grad} - \upi/2$, as shown in Appendix \ref{app:acorr}.

The characteristic amplitude of oscillation for the interfacial concentration gradient may be estimated from (\ref{eq:cgrad_wall}),
\begin{equation}
    O\left(\cgrad \right) \sim \varepsilon (1 +\bar a \sqrt\Omega) = \delta_\mathit{grad}.
    \end{equation}
Equation (\ref{eq:acorr_def}) may be rewritten in terms of the different oscillation amplitudes about the equilibrium radius $\bar a$, 
\begin{equation}
    \bar a + \delta \approx (\bar a + \delta_p) (\bar a + \delta_\mathit{corr})/\bar a.
\end{equation}
Linearization provides
$\delta =  \delta_p + \deltacorr + O(\delta_p \deltacorr)$.
Amplitudes $\delta_p$ and $\deltacorr$ may be reasonably approximated as the product of the characteristic time derivative and the characteristic timescale, $1/\Omega$. We obtain
\begin{equation}
    \frac{1}{\Omega} \frac{\dd a_p}{\dd \tau} 
    \sim \frac{\varepsilon \bar a}{3} = \delta_p 
\end{equation}
which is consistent with (\ref{eq:ap_def}). Likewise, making use of (\ref{eq:dacorr}),
\begin{equation}
     \frac{1}{\Omega}\frac{\dd \acorr}{\dd \tau} 
    \sim \frac{\Lambda \varepsilon}{\bar a\Omega}\left(1 + \bar a\sqrt\Omega\right)
    = \delta_\mathit{corr}.
\end{equation}

\begin{figure}
  \centerline{\includegraphics[width= 0.9\textwidth]{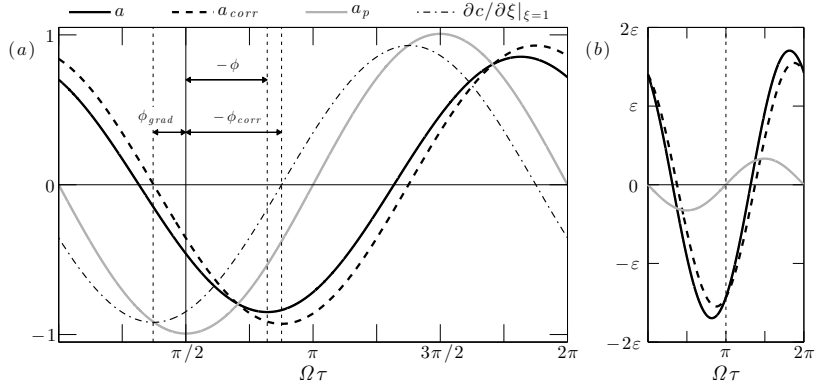}}
  \caption{Numerically obtained oscillation waveforms for $\Omega \!=\! 1$, with $\bar a \!=\! 1$, $\varepsilon \!=\! 0.01$ and $\Lambda \!=\! 0.828$. Phase shifts are annotated in plot (\textit{a}) showing the normalized waveforms for strains $(a \!-\! \bar a)/\delta$, $(\acorr \!-\! \bar a)/\delta_\mathit{corr}$, $(a_p \!-\! \bar a_p)/\delta_p$,  and for the interfacial concentration gradient $(\partial c/\partial\xi|_{\xi \!=\! 1})/\delta_\mathit{grad}$. (\textit{b}) Unscaled strain waveforms $a/\bar a \!-\!1$, $\acorr/\bar a \!-\!1$ and $a_p/\bar a \!-\!1$.}
  \label{fig:waveform_omega1}
\end{figure}

Two limiting cases arise. 
For small $\Omega \ll 1$,  we see that $\delta  \approx \deltacorr \sim \varepsilon\Lambda/\bar a \Omega$. The bubble oscillation amplitudes are provided entirely by mass transfer across the bubble surface, the pressure-induced expansion and contraction amplitude is negligible in comparison. The behaviour of the different phase shifts as $\Omega$ approaches this limit is: 
\begin{equation}
    \mbox {as} \quad \Omega \rightarrow 0, 
    \quad \phi_\mathit{grad} \rightarrow  0, \ \ 
    \phi_\mathit{corr} \rightarrow  -\upi/2, \ \
    \phi \rightarrow  \phi_\mathit{corr} \rightarrow -\upi/2.
\end{equation}
Taking $\Omega = 1$, $\bar a = 1$ (figure \ref{fig:waveform_omega1}), the dominant contribution in $\delta$ and hence $\phi$ still come from $\deltacorr$ and $\phi_\mathit{corr}$. The relative contributions of $\deltacorr$ and $\delta_p$ may be estimated from $\deltacorr /\delta_p \sim 6\Lambda \approx 5$.

\begin{figure}
  \centerline{\includegraphics[width= 0.9\textwidth]{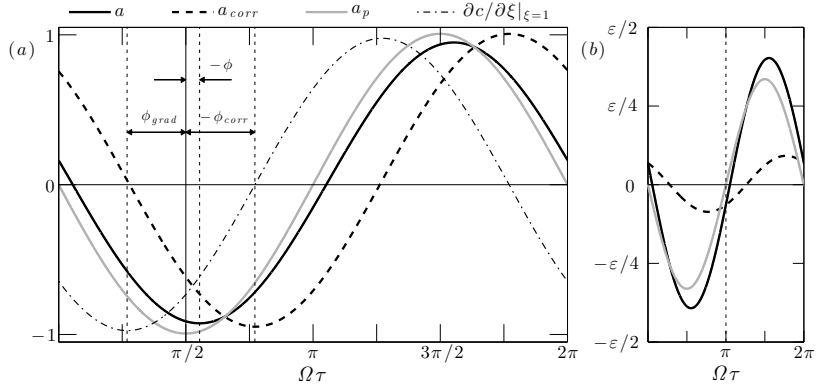}}
  \caption{Oscillation waveforms for $\Omega \!=\! 100$, with $\bar a \!=\! 1$, $\varepsilon \!=\! 0.01$ (see caption of figure \ref{fig:waveform_omega1}).}
  \label{fig:waveform_omega100}
\end{figure}

For large $\Omega \gg 1$,  we see that $\delta  \approx \delta_p \sim \varepsilon \bar a /3$. There is negligible mass transfer during an individual oscillation due to the short oscillation period. The phase shifts behave according to: 
\begin{equation}
    \mbox {as} \quad \Omega \rightarrow \infty, 
    \quad \phi_\mathit{grad} \rightarrow  \upi/4, \ \
    \phi_\mathit{corr} \rightarrow  -\upi/4, \ \
    \phi \rightarrow  \phi_p = 0.
\end{equation}
Taking $\Omega = 100$, $\bar a = 1$ (figure \ref{fig:waveform_omega100}), $a_p$ now provides the main contribution since $\deltacorr/\delta_p \sim 3\Lambda/10 \approx 1/4$.

\begin{figure}
  \centerline{\includegraphics[width= 0.9\textwidth]{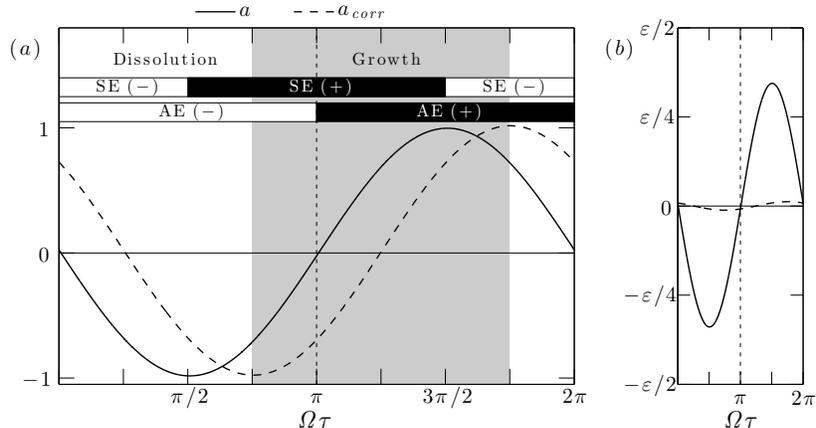}}
  \caption{Strain waveforms for $\Omega \!=\! 5000$, with $\bar a \!=\! 1$, $\varepsilon \!=\! 0.01$ and $\Lambda \!=\! 0.828$. (a) Normalized strain waveforms  $(a \!-\! \bar a)/\delta$ and $(\acorr \!-\! \bar a)/\delta_\mathit{corr}$. SE and AE refer to the `shell effect' and `area effect', which may increase ($+$) or hinder ($-$) mass transfer across the bubble, be it growth ($\dd\acorr/\dd\tau\!>\!0$, grey background) or dissolution. (\textit{b}) Unscaled strain waveforms $a/\bar a \!-\!1$ and $\acorr/\bar a \!-\!1$.}
\label{fig:AESE}
\end{figure}

Ultimately, the difference in the phase lags $\phi$ and $\phi_\mathit{corr}$ partly induced by the history effect has a direct impact on the relative contributions of the area effect and the so-called shell effect \citep{Eller&Flynn:1965}, the two driving mechanisms behind bubble growth in rectified diffusion. 
The area effect refers to the increase of the bubble interfacial area during expansion and the subsequent decrease during contraction. 
The velocity field generated by the bubble oscillations also influences the gas transport through the advection term in the advection-diffusion equation. This influence is referred to as the shell effect. More specifically, we can regard the shell effect as the advection-induced squeezing of the concentration boundary layer when the bubble radius expands and similarly the stretching when the radius contracts. In turn, this effect yields either a steeper or shallower concentration gradient at the bubble surface compared to that of a motionless bubble of the same size. As a result, the area effect increases mass transfer with respect to the motionless equilibrium bubble when $a>\bar a$. Similarly, the shell effect has an amplifying effect on the interfacial mass flux when $\dd a/\dd\tau>0$.

Let us consider the high frequency limit where $\phi = 0$, $\phi_\mathit{corr} = \upi/4$ and $a \approx a_p$. The area and shell effects act non-symmetrically in the bubble dissolution ($\dd \acorr/\dd\tau<0$) and growth ($\dd \acorr/\dd\tau>0$) periods, which are of equal duration in time.
Figure \ref{fig:AESE} shows that the area and shell effect increase mass transfer during 3/4 of the growth period and 1/4 of the dissolution period. Consequently, these effects reduce mass transfer during 1/4 of the growth period and 3/4 of the dissolution period. As a result, bubble growth is always promoted.

%%%%%%%%%%%%%%%%%%%%%%%%%%%%%%%%%%%%%%%%%%%%%%
%% SECTION: Conclusions
%%%%%%%%%%%%%%%%%%%%%%%%%%%%%%%%%%%%%%%%%%%%%%
\section{Conclusions}
\label{sec:conclusions}

The contribution of any past mass transfer events between a gas bubble and its liquid surroundings towards the current diffusion-driven bubble growth or dissolution dynamics has been referred to as the `history effect'. The history effect arises from the non-instantaneous development of the concentration boundary layer in response to changes in the concentration at the bubble interface caused, for instance, by variations of the ambient pressure in time. Put in another way, the current state of the concentration profile must be naturally conditioned by the preceding time-history of the concentration field itself. 
As a consequence, the mass flux across the bubble is conditioned by the preceding time-history of the boundary condition. This very last notion is the essence of the history effect. It has been shown that the contribution of the history effect in the current interfacial concentration gradient is fully contained within a memory integral of the interface concentration.

Under the assumption that advection effects are small, the integral term has been used to derive the governing equation for bubble dynamics concerning the canonical case of a spherical bubble suspended in a quiescent liquid. It has been termed as the Epstein-Plesset with history term (EPH) equation. This equation is not restricted to a constant pressure-time history and does not make use of the quasi-static radius approximation. It is however expressed in nonlinear time $\ttau$, which can be related to the standard time once the EPH equation has been solved for $a(\ttau)$, the time evolution of the bubble radius.

The EPH equation has been analytically solved for the case of multiple step-like jumps in pressure. The nature and relevance of the history effect in the bubble dynamics has been assessed through illustrative examples. A future companion paper shall deal with the experimental  and numerical analysis of the the history effect regarding a sessile bubble exposed to such step-like pressure-time histories. 

Finally, we have investigated the role of the history effect concerning the problem of mass transfer across a non-inertial bubble that pulsates under harmonic pressure forcing. The history effect has been shown to induce a phase shift in the interfacial concentration gradient with respect to the phase of the interfacial concentration. At sufficiently high forcing frequencies, the oscillatory mass flux across the bubble is entirely provided by the history integral term and the aforementioned phase shift asymptotically tends to $\upi/4$. This phase shift causes the shell effect and area effect (the two driving mechanisms behind rectified diffusion) to act non-symmetrically on growth and dissolution during each individual bubble oscillation period. Growth is always favoured as a result.

%%%%%%%%%%%%%%%%%%%%%%%%%%%%%%%%%%%%%%%%%%%%%%
%% Acknowledgements
%%%%%%%%%%%%%%%%%%%%%%%%%%%%%%%%%%%%%%%%%%%%%%
%\acknowledgements
The authors acknowledge the support of the Spanish Ministry of Economy and Competitiveness through grant DPI2014-59292-C3-1-P, partly financed by European funds. The authors are deeply grateful to A. Moreno-Soto especially for his experimental guidance during the acquisition of figure \ref{fig:intro}. The authors are also grateful to O. Enr\'iquez for promoting this study and D. Lohse for his helpful suggestions.

%%%%%%%%%%%%%%%%%%%%%%%%%%%%%%%%%%%%%%%%%%%%%%
%% APPENDICES
%%%%%%%%%%%%%%%%%%%%%%%%%%%%%%%%%%%%%%%%%%%%%% 
\appendix
%%%%%%%%%%%%%%%%%%%%%%%%%%%%%%%%%%%%%%%%%%%%%%
%% APPENDIX: Derivation of the history integral term
%%%%%%%%%%%%%%%%%%%%%%%%%%%%%%%%%%%%%%%%%%%%%%
\section{Derivation of the history integral term}
\label{app:historyterm}
Upon the substitution $f(\xi, \ttau) = \xi \:c(\xi, \ttau)$,
and furthermore neglecting any arising $(\dd a /\dd \ttau)/a$ terms,
Equation (\ref{eq:diff}) transforms to
\begin{equation} \label{eq:f_eq}
   \frac{\partial f}{\partial \ttau} = \frac{\partial^2 f}{\partial \xi^2}.
\end{equation}
Taking the Laplace transform of (\ref{eq:f_eq}) subject to the initial condition $f(\xi > 1, 0) = 0$ yields
\begin{equation}
  s\hat f = \frac{\dd^2 \hat f}{\dd \xi^2}.
\end{equation}
A solution compatible with boundary conditions $f(1, \tau) = c_s(\ttau)$ and $f(\infty, \ttau) = 0$ is found to be
\begin{equation} \label{eq:fhat}
  \hat f(\xi;s) = \hat c_s \mathrm{e}^{-\sqrt s(\xi-1)},
\end{equation}
where $\hat{c}_s$ is a function of $s$, to be determined from the boundary conditions.
The transformed concentration gradient across the bubble interface is
\begin{equation} \label{eq:cgrad_hat}
    \left. \frac{\partial \hat c}{\partial \xi} \right|_{\xi = 1} = 
     \left. \frac{\partial \hat f}{\partial \xi}\right|_{\xi = 1} - \hat f(1;s)
     = -(1+\sqrt s)\hat c_s .
\end{equation}
The initial value of $c_s$ is required. It is given by 
\begin{equation}
    c_s(0) = p_0+ \sigma/a_0 -\Upsilon = c_{s0},
\end{equation}
where $p_0 = p(0)$ and $a_0 = a(0)$ denote the initial pressure and radius respectively.
Note that in perfect initial saturation conditions, the surface concentration is the same as the gas concentration in the liquid, $C_s(0) = C_\infty$. This amounts to $\Upsilon = p_0$ and $c_{s0} = \sigma/a_0>0$, corresponding to slow dissolution purely driven by the Laplace pressure. Taking then the inverse Laplace transform of (\ref{eq:cgrad_hat}) results in
\begin{equation}
    -\left.\frac{\partial c}{\partial \xi}\right|_{\xi = 1} =  c_s
     +\int_{0}^{\ttau} \frac{1}{\sqrtsign{\upi(\ttau-\tx)}}
     \left[ \frac{\dd c_s}{\dd \tx} + c_{s0}\delta( \tx) \right] \:\dd \tx .
    \end{equation}
The term containing the Dirac delta $\delta(\tx)$ may be directly integrated. This finally sets the concentration gradient at the interface to be 
\begin{equation} 
    -\left.\frac{\partial c}{\partial \xi}\right|_{\xi = 1} = c_s
     +\frac{c_{s0}}{\sqrt{\upi\ttau}} 
     + \int_{0}^{\ttau} \frac{1}{\sqrtsign{\upi(\ttau- \tx)}}
     \frac{\dd c_s}{\dd \tx} \:\dd \tx.
\end{equation}
It is not surprising that \cite{Jones&Zuber:1978} obtained an analogous expression for the heat flux across a vapour bubble. In fact, the history integral provides the general solution (for the spatial derivative at the boundary) for the canonical diffusion problem consisting of a scalar field $U(z,t)$ obeying
\begin{equation}
\frac{\partial U}{\partial t} = \alpha \frac{\partial^2 U}{\partial z^2},
\end{equation}
with boundary and initial conditions $U(0, t) = U_s(t)$ and $U(z,0) = U(\infty, t) = 0$.  The solution for the gradient at the boundary $z = 0$ \citep[see, for example,][]{Landau&Lifshitz:1987A} reads
\begin{equation}
-\left. \frac{\partial U}{\partial z}\right|_{z=0} = 
\frac{1}{\sqrtsign{\alpha}} \int_{0}^{t} \frac{1}{\sqrtsign{\upi(t-x)}}
     \frac{\dd U_s}{\dd x} \:\dd x. 
\end{equation}
The history integral is therefore essential in determining the heat flux across a plate or the shear stress exerted by a viscous fluid on a plate moving in its plane (often referred to as the Basset force). While the history integral has been identified to play an important role in the rectilinear motion of bubbles in viscous flows \citep{Magnaudet&Legendre:1998}, its effect has been overlooked in diffusion-driven bubble dissolution and growth.

%%%%%%%%%%%%%%%%%%%%%%%%%%%%%%%%%%%%%%%%%%%%%%
%% APPENDIX: Solutions
%%%%%%%%%%%%%%%%%%%%%%%%%%%%%%%%%%%%%%%%%%%%%%
\section{Solutions for step changes in pressure} 
\label{app:solution_steps}
\subsection{Analytical solution}
Let $p_n$ be the pressure at the  $n$-th plateau (constant pressure segment) after jump $n$. Its value is thus
\begin{equation}
    p_n = p_0+ \sum_{j = 1}^n \Delta p_j  
    \quad \mbox{for} \quad \tT_n^+ \leq \ttau \leq \tT_{n+1}^-. 
\end{equation}
%and shall be used alongside the limiting conditions $\ttau_0 = 0$, $\Delta p_0 = 0$ and $\ttau_{N+1} = \infty$. 
Time coefficient $\tT_n^-$ describes the instant in time right before the $n$-th pressure jump at $\ttau = \tT_n$, while $\tT_n^+$ refers to the moment right after.

\subsubsection*{Solution for the Epstein-Plesset with history term equation}
The EPH equation (\ref{eq:EPH_nost}) may be integrated in time within the $n$-th segment as
\begin{eqnarray} 
\int^{a(\ttau)}_{a(\tT_n^-)} \: \dd \ln a + \frac 13 &&\int^{p_n}_{p_{n-1}} \: \dd \ln p
= -\frac{\Lambda}{p_n} \left[ 
\int^{\ttau}_{\tT_n^+} \left(p_n - \Upsilon + \frac{p_0 - \Upsilon}{\sqrt{\upi \ty}} \right)\: \dd \ty
\right. \nonumber\\
&&\left. + \int^{\ttau}_{\tT_n^+} \left\{ \int^{\ty}_{0} \frac{1}{\sqrt{\upi(\ty-\tx)}}
\sum^n_{j= 1} \Delta p_j \delta(\tx-\tT_j) \: \dd \tx \right\} \: \dd \ty
\right].
\end{eqnarray}
Evaluating this integral finally yields, for segments $n = 1$ to $n = N$,
\begin{eqnarray} \label{eq:anaEPH1} 
a(\ttau) &=& a(\tT_n^-) \left(\frac{p_n}{p_{n-1}} \right)^{-1/3}
\exp \left\{-\frac{\Lambda}{p_n} \left[(\ttau- \tT_n)(p_n-\Upsilon) 
+ \frac{2}{\sqrt{\upi}} (p_0-\Upsilon)\left(\sqrt{\ttau} -\sqrt{\tT_n}\right) \right. \right. \nonumber\\
&& \left. \left. + \mbox{ } \frac{2}{\sqrt{\upi}} \sum_{j = 1}^n \Delta p_j 
\left (\sqrt{\ttau - \tT_j} - \sqrt {\tT_n - \tT_j} \right) \right]
\right\}
 \quad \mbox{for} \quad \tT_n^+ \leq \ttau \leq \tT_{n+1}^-.
\end{eqnarray}
The initial condition is $a(0) = a_0$. 
The end radius from any segment $a(\tT_n^-)$ must be first computed before moving on to the next. We begin with segment $n = 0$, i.e. before the first pressure jump. The radius dynamics in this initial segment are evidently identical to the solution in (\ref{eq:epn-sol}) and are straightfowardly given by
\begin{equation} \label{eq:anaEPH2}
a(\ttau) = a_0
\exp \left\{-\Lambda (p_0-\Upsilon) \left( \ttau + 2\sqrtsign{\ttau/\upi} \right)
\right\}
 \quad \mbox{for} \quad 0 \leq \ttau \leq T_1^-.
\end{equation}

\subsubsection*{Solution for the concentration field}

From (\ref{eq:fhat}), the Laplace transformed concentration field $\hat c(\xi;s)$ may be conveniently split as the product of two separate functions $\hat f(s)$ and $\hat g(\xi;s)$ as follows:
\begin{equation}
\hat c(\xi;s) = \hat g(s) \times \hat h(\xi;s) = s \hat c_s \times \frac{\mathrm{e}^{-\sqrt s (\xi -1)}}{\xi s}.
\end{equation}
The inverse Laplace transforms of these two functions may be shown to be
\refstepcounter{equation}
$$
g(\ttau) = \frac{\dd c_s}{\dd \tau} + c_{s0}\delta(\ttau), \qquad 
h(\ttau) = \frac{1}{\xi}\mathrm{erfc}\left( \frac{\xi-1}{2 \sqrtsign{\ttau}}\right) .
    \eqno{(\theequation{\mathit{a},\mathit{b}})}
$$
Using these results, the concentration field in the time domain $c( \xi, \ttau)$ may be then computed by means of the convolution theorem. This gives
\begin{equation} \label{eq:cfield}
    c(\xi,\ttau) = \frac{c_{s0}}{\xi} \mathrm{erfc}\left(\frac{\xi-1}{2\sqrtsign{\ttau}}\right)
    + \frac{1}{\xi}\int_{0}^{\ttau} \frac{\dd c_s}{\dd \tx} 
    \mathrm{erfc}\left(\frac{\xi-1}{2\sqrtsign{\ttau- \tx}}\right) \:\dd \tx. 
\end{equation}
Surface tension is to be neglected once again. This renders $c_{s0} = p_0-\Upsilon$ and $\dd c_s/\dd \ttau = \dd p/\dd \ttau$, the latter modelled in (\ref{eq:pstep}). After inserting these expressions into (\ref{eq:cfield}), the time evolution of the concentration field at every $n$-th segment is finally obtained:
\begin{equation}
    c(\xi,\ttau) = 
    \left\{ \begin{array}{ll}
         \displaystyle \frac{p_0-\Upsilon}{\xi} \mathrm{erfc}\left(\frac{\xi-1}{2\sqrtsign{\ttau}}\right)
         & \mbox{for} \quad 0 \leq \ttau \leq T_1^-, \\[8pt]
         \displaystyle \frac{p_0-\Upsilon}{\xi} \mathrm{erfc}\left(\frac{\xi-1}{2\sqrtsign{\ttau}}\right)
         +\frac{1}{\xi} \sum_{j = 1}^n \Delta p_j \: \mathrm{erfc}\left(\frac{\xi-1}{2\sqrtsign{\ttau- \tT_j}}\right)
          & \mbox{for} \quad \tT_n^+ \leq \ttau \leq \tT_{n+1}^-.
 \end{array} \right.
\end{equation}

%% SUBSECTION: Advection and surface tension effects
\subsection{Numerical model}
\label{app:numerical_model}
The pressure step functions are analytically modelled by the logistic function
\begin{equation} \label{eq:pttau}
    p(\ttau) = p_0 + \frac{1}{2}\sum_{n=1}^N{ 
            \Delta p_n \left\{ 1 +\tanh[\tilde k (\ttau-\tT_n)]\right\}},
\end{equation}
where $\tilde k$ is a large constant. The Heaviside functions are recovered as $\tilde k \rightarrow \infty$, whilst here the value $\tilde k = 1000$ was deemed as sufficiently large.
The time derivative of the pressure reads 
\begin{equation} \label{eq:p'ttau}
    \frac{\dd p}{\dd \ttau} = \frac{\tilde k}{2}\sum_{n=1}^N{\Delta p_n \mathrm{sech}^2[\tilde k (\ttau-\tT_n)]}.
\end{equation}
The governing equations will be numerically integrated in nonlinear time $\ttau$, purely for consistency with the analytical derivation. However, we shall conveniently establish the pressure-time history input and present the bubble size-history in the linear time $\tau$. To this end, the physical time $\tau$ was computed at each timestep by integrating $\dd \tau = a^2\dd\ttau$. 
Provided $\tilde k$ is large enough,  (\ref{eq:pttau}) and (\ref{eq:p'ttau}) may be computed from
\refstepcounter{equation}
$$
    p(\ttau- \tT_n) = p(\tau-T_n) = p_0 + \frac{1}{2}\sum_{n=1}^N{ 
            \Delta p_n \left\{ 1 +\tanh[k (\tau-T_n)]\right\}}, \quad
    \frac{\dd p}{\dd \ttau} = a^2 \frac{\dd p}{\dd \tau},
    \eqno{(\theequation{\mathit{a},\mathit{b}})} \label{eq:ptau}
$$
where use of $k = \tilde k$ has been made. Note that the precise value of $k$ is not relevant provided it is large. The radius dynamics can then be obtained by integrating (\ref{eq:mconsdim}), namely
\begin{equation} \label{eq:a'sim}
    \frac{\dd a}{\dd \ttau} =  a \left( \Lambda\left.\frac{\partial c}{\partial \xi}\right|_{\xi = 1}
        - \frac 13 \frac{\dd p}{\dd \ttau} \right)\left(p + \frac{ 2\sigma}{3a}\right)^{-1},  
\end{equation}
subject to the prescribed initial size $a_0$.
The concentration gradient at the interface is solved for numerically from (\ref{eq:ADdim}) using a finite-differences scheme. Equation (\ref{eq:ADdim}) is written here again for the reader's convenience: 
\begin{equation} \label{eq:ADdim2}
    \frac{\partial c}{\partial \ttau} 
    + \Pen(\ttau) \left(\frac{1}{\xi^2}-\xi \right) \frac{\partial c}{\partial \xi} 
    =  \frac{1}{\xi^2} \frac{\partial}{\partial \xi} \left(\xi^2
\frac{\partial c}{\partial \xi} \right),
\end{equation}
where $\Pen(\ttau) = (\dd a/\dd \ttau)/a$.
The required boundary and initial conditions are
\refstepcounter{equation}
$$
    c(1,\ttau) = p(\ttau) + \sigma/a - \Upsilon, \qquad
     \left.\frac{\partial c}{\partial \xi}\right|_{\xi = \infty} = 0, \qquad
     c(\xi>1,0) = 0.
    \eqno{(\theequation{\mathit{a-}\mathit{c}})} 
    \label{eq:boundary_conditions_numerical}
$$

%%%%%%%%%%%%%%%%%%%
\section{Conditions for small amplitude, isothermal oscillations}
\label{app:conditions}

Following the work of \cite{Prosperetti:1977} (see figure 1 of that paper), the assumption of isothermal oscillations (polytropic exponent equal to unity) may be safely assumed, provided the thermal P\'eclet number based on the oscillation frequency is smaller than one: $\Omega_\mathit{th} \equiv 2 \upi f_c R_c^2/D_\mathit{th} < 1$, with $D_\mathit{th} = k_g/\rho_g c_{v,g}$. Symbols $k_g$, $\rho_g$ and $c_{v,g}$ denote the gas thermal conductivity, density and specific heat at constant volume respectively. Consequently, $\Omega$ does not need to be small for this assumption to hold, as long as $\Omega < D_\mathit{th}/D_m$. As an example, for air at 300 K, $D_\mathit{th}/D_m \sim 10^4$.

Assumption (a) additionally implies that the oscillation frequency $f_c$ must be  much smaller than the resonance frequency of the bubble.
The low frequency limit, for history effects to be visible, is given by $f_c \gtrsim 1/t_m$, where $t_m = R_c^2/D_m$ is the timescale of mass transfer by diffusion. 
When $f_c \ll 1/t_m$, the rate of change of the bubble interfacial concentration is very slow compared to mass diffusion. The concentration field has enough time to reach the quasi-steady solution characterized by the absence (full dissipation) of the history effect.
Summarizing these last ideas, $\Omega$ must additionally satisfy
\begin{equation}
1 \lesssim \Omega \ll \Omega_\mathit{res},
\end{equation}
where $\Omega_\mathit{res} = \sqrt{3 P_c/\rho_l} R_c/D_m$, is the P\'eclet number based on Minnaert's resonance frequency.
%(\ref{eq:omega_range}b).
As an example, consider bubbles trapped in magma chambers exposed to earthquake-induced shaking with typical periods of 1--20 s \citep{Ichihara&Brodsky:2006}. This results in $\Omega \sim 1$--$100$ for a 100 $\umu$m bubble. 
 %($R_c \sim 100 \ \mathrm{\mu m}$, $D_m \sim 10^{-9} \ \mathrm{m^2s^{-1}})$. 
 Likewise, a 10 $\umu$m bubble in the tissue of a marine mammal exposed to 5--5000 Hz sonar \citep{Crum&Mao:1996} corresponds to $\Omega \sim 30$--$3000$.

Except for small values of $\Omega \ll 1$ outside our range of interest, (where the oscillation period is slow enough so that gas diffusion across the bubble surface results in large volumetric oscillations), $\delta$ will always be of the same order of magnitude as the pressure-induced isothermal expansion and contraction amplitude. Comparison of (\ref{eq:a_def}) and (\ref{eq:ap_def}) gives $\delta/\bar a \sim \varepsilon/3$. Therefore, (b) requires that $ \varepsilon \ll 1$.

Turning now to (c), consider a bubble in a gas-saturated liquid $\Upsilon =1$, with negligible Laplace pressure $\sigma = 0$. Under these conditions, the growth rate purely due to rectified diffusion is expected to be largest. For such a case, the asymptotic growth rate of $\bar a$ due to rectified diffusion under assumptions (a)--(c) was found to be reasonably well approximated by \cite{Hsieh&Plesset:1961},
\begin{equation} \label{eq:hp}
%\frac{\dd \bar a}{\dd \tau} =  \frac 23 \Lambda\varepsilon^2. 
\dd \bar a/\dd \tau =   (2/3) \Lambda\varepsilon^2.
\end{equation}
The inverse of the characteristic timescale for growth is then $\tau_b^{-1} \sim \Lambda \varepsilon^2/\bar a$, which must be much smaller than  $\Omega$.
Let us now consider low forcing frequencies in addition to a highly supersaturated (say $\Upsilon \sim 2$--3) or undersaturated ($\Upsilon \approx 0$) liquid together with a non-negligible Laplace pressure. The likely existence of high diffusion rates will result in fast diffusion-driven bubble growth or dissolution. Consequently, the equilibrium radius $\bar a$ may no longer be constant over an individual oscillation period \citep{Ilinskii:2008}, i.e. assumption (c) no longer holds.
The inverse of the characteristic time for bubble growth or dissolution may be estimated from (\ref{eq:ep}) as
$\tau_b^{-1} \sim |\Upsilon-1 -\sigma/\bar a|\Lambda /\bar a^2$.
Thus, (c) imposes that the solution saturation level, pressure amplitude and frequency must fulfil the following inequalities:
\refstepcounter{equation}
$$
    |\Upsilon-1 -\sigma/\bar a| \ll \Omega \bar a^2/\Lambda, \qquad
    \varepsilon^2 \ll  \bar a \Omega/ \Lambda.
    \eqno{(\theequation{\mathit{a},\mathit{b}})}
$$
Finally note that if surface tension and undersaturation are to be overcome, equating Hsieh-Plesset solution (\ref{eq:hp}) with the Epstein-Pesset solution (\ref{eq:ep}) leads to an approximate, simplistic threshold condition for growth \citep{Safar:1968}:  $\varepsilon^2  =  (3/2) \left( 1-\Upsilon + \sigma/\bar a \right)$. As noted by \cite{Safar:1968}, this threshold is only valid for large isothermal bubbles in a liquid close to saturation ($\Upsilon \approx 1$) under sufficiently high ambient pressures such that the Laplace pressure is comparatively small: $\sigma/\bar a \ll 1$. 
As an example, 
$\varepsilon \sim 0.01$ would be required to overcome the surface tension driven dissolution of a CO$_2$ bubble of size $R_c \sim 100$ $\umu$m in saturated water at $P_c \sim 20$ MPa.

\section{The oscillatory problem} 
\label{app:oscillatory} 
Valuable insight on the nature of the concentration field may be gained by performing an order of magnitude analysis on the governing mass transport equation:
\begin{equation} \label{eq:ad_dim2}
\frac{\partial c}{\partial \tau} +\frac{\Pen(\tau)}{a^2} \left( \frac{1}{\xi^2}-\xi \right) \frac{\partial c}{\partial \xi}  = \frac{1}{a^2 \xi^2}\frac{\partial}{\partial \xi}\left(\xi^2 \frac{\partial c}{\partial \xi} \right) 
\end{equation} 
where $\Pen(\tau) = (\dd a/\dd \tau) \: a = R \dot R/D_m$. Its magnitude is given by $O(\Pen) \sim \varepsilon\Omega\bar a^2/3$ since
$O(a) \sim \bar a$ and
$O(\dd a/\dd \tau) =  O(\dd a_p/\dd \tau) \sim \varepsilon\Omega\bar a/3$.
Bear in mind that $\bar a$ will often be of order unity (provided $R(t)$ is comparable to the chosen characteristic radius $R_c$), but we shall carry the analysis allowing for any magnitude of $\bar a$. 

%Advection term amplifies the interfacial concentration gradient but does not change  its phase. 

Evaluation of (\ref{eq:ad_dim2}) on $\xi = 1$ makes the advection term become identically zero. The diffusive term must therefore be of leading order in a layer bounded by $1<\xi < 1+l$ where $O(\partial c) = O(\Delta c_s) \sim \varepsilon$. Parameter $l$ is the dimensionless penetration depth of diffusion using $R(t)$ as the lengthscale, i.e. $l = L/R(t)$.  The relevant lengthscales are $O(\xi) \sim 1$ and  $O(\partial\xi) \sim l$, while the characteristic timescale per oscillation is $O(\partial\tau) \sim \Omega^{-1}$. The magnitudes of the unsteady and diffusive terms in (\ref{eq:ad_dim2}) are given by
\refstepcounter{equation}
$$
    O\left(\frac{\partial c}{\partial \tau}\right) \sim \varepsilon \Omega, \qquad
    O\left(\frac{1}{a^2 \xi^2}\frac{\partial}{\partial \xi}\left(\xi^2 \frac{\partial c}{\partial \xi} \right) \right) \sim \frac{\varepsilon}{\bar a^2 l^2}.
    \eqno{(\theequation{\mathit{a},\mathit{b}})}
$$
Balancing these two terms yields $l \sim 1/(\bar a \sqrt\Omega)$. Note that the equivalent dimensional penetration depth is $L \sim \sqrtsign{D_m/(2\upi f_c)}$, and it is independent of the bubble size.
In the low frequency limit, it follows from the Epstein-Plesset solution that $l \sim 1$. Hence, $(\bar a^2\Omega)^{-1} \sim 1$ is of the order of the diffusion timescale associated to a lengthscale equal to the current bubble size. Lengthscales are now $O(\xi) = O(\partial \xi) \sim 1$ and it is easy to show that the advection term scales as $\sim \varepsilon^2\Omega$.
Approaching the high frequency limit, the penetration depth is much smaller than current bubble radius ($l \ll 1$), i.e. when  $\bar a^2 \Omega \gg 1$. A series expansion of the advection term taking $\xi = 1+l$ reveals that the magnitude of the dimensionless advection term is independent of $l$ since: 
\begin{equation}
O\left(\frac{\Pen(\tau)}{a^2} \:\left( \frac{1}{\xi^2}-\xi \right) \:\frac{\partial c}{\partial \xi}\right) 
\sim \frac{\varepsilon\Omega}{3} \:(3l) \:\frac{\varepsilon}{l}
\sim  \varepsilon^2\Omega. 
\end{equation}
We conclude that under the small amplitude oscillation restriction, the magnitude of the dimensionless advection term in (\ref{eq:ad_dim2}) is always smaller than the unsteady and diffusive terms by a factor of $\varepsilon \ll 1$. Therefore, the dimensionless advection term may be neglected in the oscillatory problem.
Moreover, following the discussion presented in section \ref{sec:oscillatory_problem}, we may assume $a(\tau) = \bar a(\tau) + O(\delta)  \approx \bar a$ to remain constant over an oscillation period and additionally take the liquid to be saturated: $\Upsilon = 1$.

Hence, letting $f = \xi\: c$, Equation (\ref{eq:ad_dim2}) is reduced to a parabolic equation,
\begin{equation} \label{eq:f_eqosc}
    \frac{\partial f}{\partial \tau}  = \frac{1}{\bar a^2}\frac{\partial^2 f}{\partial \xi^2}
\end{equation}
together with boundary conditions
\refstepcounter{equation}
$$
c(1,\tau) = f(1,\tau) = c_s(\tau) = \varepsilon\sin\Omega\tau, \qquad 
c(\infty,\tau) = f(\infty,\tau) = 0.
    \eqno{(\theequation{\mathit{a},\mathit{b}})}
    \label{eq:bcs_fosc}
$$
Resorting to the same treatment given to (\ref{eq:f_eq}) involving Laplace transforms, we arrive from (\ref{eq:f_eqosc}) to an analogous expression for (\ref{eq:cgrad}), the concentration gradient evaluated at the interface:
\begin{equation} 
    -\left.\frac{\partial c}{\partial \xi}\right|_{\xi = 1} = c_s
     +\frac{c_{s0}\:\bar a}{\sqrt{\upi\tau} } 
     + \bar a \int_{0}^{\tau} \frac{1}{\sqrtsign{\upi(\tau- x)}}
     \frac{\dd c_s}{\dd x} \:\dd x.
\end{equation}
Under our particular conditions, $c_{s0} =0$. The integral term, with $c_s(\tau) = \varepsilon\sin\Omega\tau$, may be evaluated from an identity provided by \cite{Stepanyants&Yeoh:2009}, yielding
\begin{equation} \label{eq:cgrad_int}
    -\left.\frac{\partial c}{\partial \xi}\right|_{\xi = 1} = \varepsilon\sin\Omega\tau 
     + \bar a\varepsilon\Omega \int_{0}^{\tau} \frac{\cos\Omega x}{\sqrtsign{\upi(\tau- x)}} \:\dd x
     = \varepsilon \left[\sin\Omega\tau + \bar a \Omega \:\Real\left\{\frac{\mathrm{e}^{\mathrm{i}\Omega\tau} \:\mathrm{erf}\sqrt{\mathrm{i}\Omega\tau}}{\sqrt{\mathrm{i}\Omega}}\right\} \right].
 \end{equation}
The asymptotic expansion as $\tau \rightarrow \infty$ (since we are interested in the steady periodic state solution) of the last term is
\begin{equation}
%\lim_{\tau\rightarrow \infty} \Real\left\{\frac{\mathrm{e}^{\mathrm{i}\Omega\tau} \:\mathrm{erf}\sqrt{\mathrm{i}\Omega\tau}}{\sqrt{\mathrm{i}\Omega}}\right\}
\Real\left\{\frac{\mathrm{e}^{\mathrm{i}\Omega\tau} \:\mathrm{erf}\sqrt{\mathrm{i}\Omega\tau}}{\sqrt{\mathrm{i}\Omega}}\right\}
\sim \frac{1}{\sqrt{2\Omega}}\left(\cos\Omega\tau + \sin\Omega\tau\right).
\end{equation}
Finally, inserting this result into (\ref{eq:cgrad_int}) we obtain
\begin{equation} \label{eq:cgrad_wall2}
    \left.\frac{\partial c}{\partial \xi}\right|_{\xi = 1} 
     = -\varepsilon \left[\sin\Omega\tau + {\bar a}\sqrt\Omega\sin\left(\Omega\tau + \frac{\upi}{4}\right) \right].
 \end{equation}
The last term is hence the contribution from the history integral.
As reiterated by \cite{Fyrillas&Szeri:1994}, the problem of the oscillating bubble is completely analogous to Stokes' second problem of viscous flow near an oscillating flat plate \citep[see for example][]{Landau&Lifshitz:1987A}. The concentration profile must be a harmonic function in $\tau$ with the same frequency as the boundary condition $c_s(\tau) = \varepsilon\sin\Omega\tau$. Equation (\ref{eq:f_eqosc}) may be easily solved for harmonic motion considering a solution of the form 
\begin{equation}
    f(\xi, \tau) = \xi\:c(\xi,\tau) = \Imag \left\{ \hat f(\xi) \:\mathrm{e}^{\mathrm{i}\Omega\tau} \right\},
\end{equation}
compatible with the boundary conditions in (\ref{eq:bcs_fosc}). It then follows that $\hat f(1) = \varepsilon$ and $\hat f(\infty) = 0$.
The solution for the concentration profile is 
\begin{equation} \label{eq:c_profile2}
c(\xi,\tau) = \frac{\varepsilon}{\xi}\exp\left\{-\bar a\sqrtsign{\Omega/2}(\xi-1)\right\}
\sin\left\{\Omega\tau - \bar a\sqrtsign{\Omega/2}(\xi-1)\right\}
\end{equation}
which renders the following concentration gradient profile:
\begin{equation} \label{eq:cgrad_profile2}
    \frac{\partial c}{\partial \xi} = -\varepsilon \exp\left\{-\bar a\sqrtsign{\Omega/2}(\xi-1)\right\}\left[\frac{1}{\xi^2}\sin\Omega\tau + \frac{\bar a\sqrt\Omega}{\xi}\sin\left(\Omega\tau -\bar a\sqrtsign{\Omega/2}(\xi-1) +\frac{\upi}{4}\right) \right].
\end{equation}
Evaluating (\ref{eq:cgrad_profile2}) on $\xi = 1$ identically results in (\ref{eq:cgrad_wall2}).

\subsection{Treatment following \cite{Fyrillas&Szeri:1994}}
\label{app:fs}
As demonstrated by \cite{Fyrillas&Szeri:1994}, when considering a large frequency-based P\'eclet number $\Omega/(2\upi) \gg 1$, to obtain the asymptotic solution for the oscillatory concentration field one must first solve
\begin{equation}
\frac{\partial c}{\partial \hat \tau} = \frac{\partial^2 c}{\partial \eta^2}.
\end{equation}
The Lagrangian coordinate $\eta$, linear time variable $\tau'$ and nonlinear time $\hat \tau$ are defined as follows:
\begin{equation}
\eta = \left(\frac{\Omega}{2\upi}\right)^{1/2} \frac{a^3}{3} \left(\xi^3-1\right), \qquad
\tau' = f_ct = \frac{\Omega}{2\upi} \tau, \qquad 
\hat \tau(\tau') = \int^{\tau'}_0 a^4(x') \: \dd x'. 
\end{equation}
The boundary condition at interface associated to the oscillatory problem reads
\begin{equation} \label{eq:cs_fs1}
c(\eta=0, \hat \tau) = p_g(\tau') - \langle p_g(\tau')\rangle_{\hat \tau}.
\end{equation}
Neglecting inertial and viscous effects,  $p_g(\tau')$ is defined as 
\begin{equation}
p_g(\tau') = \frac{p(\tau') +\sigma/a(\tau')}{1 +\sigma/\bar a},
\end{equation}
while time averaging is computed according to
\begin{equation}
\langle f(\eta, \tau') \rangle_{\hat \tau} = \frac{1}{\hat\tau(T)} \int^{T}_0 f(\eta, \tau') a^4(\tau') \: \dd \tau'
\end{equation}
where $T$ denotes the dimensionless period of oscillation. Expressing this boundary condition as a Fourier series,
\begin{equation} \label{eq:cs_fs2}
c(\eta=0, \hat \tau) = \sum_{m=1}^\infty \left[ a_m \cos\omega_m \hat\tau + b_m \sin\omega_m\hat\tau \right], \quad \mbox{with} \quad
\omega_m = \frac{2\upi m}{\hat\tau(T)},
\end{equation}
a compatible solution is found to be
\begin{equation} \label{eq:c_fs}
c(\eta, \hat \tau) = \sum_{m=1}^\infty 
\exp\left(-\sqrt{\omega_m/2} \:\eta \right)
\left[ a_m \cos\left(\omega_m \hat\tau -\sqrt{\omega_m/2} \:\eta \right) 
+ b_m \sin\left(\omega_m \hat\tau -\sqrt{\omega_m/2} \:\eta \right) \right].
\end{equation}
The interfacial concentration gradient is thus
\begin{equation} \label{eq:cgrad_fs}
\frac{\partial c}{\partial \eta}(\eta = 0, \hat \tau) 
= -\sum_{m=1}^\infty \omega_m
\left[ a_m \cos\left(\omega_m \hat\tau + \upi/4 \right) 
+ b_m \sin\left(\omega_m \hat\tau + \upi/4 \right)  \right].
\end{equation}
Comparing (\ref{eq:c_fs}) and (\ref{eq:cgrad_fs}), it is inferred that the history effect shifts every frequency component of the (negative) interfacial concentration gradient profile with respect to the interfacial concentration by $\upi/4$ when $\Omega/(2\upi) \gg 1$.

Next, we shall prove that the solutions (\ref{eq:c_fs}) and (\ref{eq:cgrad_fs}) above will converge with those given in (\ref{eq:c_profile2}) and (\ref{eq:cgrad_profile2}) given the right set of assumptions.
We are considering harmonic pressure forcing: $p(\tau') = 1 + \varepsilon \sin 2\upi\tau'$ and $T = 1$. Making the quasi-static radius approximation, i.e. taking $ a(\tau') = \bar a + O(\delta) \approx \bar a$ as constant implies that $\hat \tau = \bar a^4 \tau'$. 
Provided the Laplace pressure is small ($\sigma/\bar a \ll 1$), then 
$p_g(\tau') \approx 1 + \varepsilon \sin 2\upi\tau' -\sigma/\bar a$ and the boundary condition (\ref{eq:cs_fs1}) becomes
\begin{equation}
c(\eta=0, \hat \tau) = \varepsilon \sin 2\upi\tau'.
\end{equation}
From (\ref{eq:cs_fs2}), we infer that $a_1 = 0$, $b_1 = \varepsilon$, $\omega_1 = 2\upi/\bar a^4$, and $a_m = b_m = 0$ for $m>1$. The solution in (\ref{eq:c_fs}) then simplifies to
\begin{equation} \label{eq:ch_fs}
c(\eta, \tau') = 
\varepsilon\exp\left(-\frac{\sqrtsign{\upi}}{\bar a^2} \:\eta \right) 
\sin\left(2\upi\tau'-\frac{\sqrtsign{\upi}}{\bar a^2} \:\eta \right). 
\end{equation}
Since $\Omega/(2\upi) \gg 1$, we have seen that the boundary layer thickness is very small in comparison with the bubble radius. We are in fact in the limit $\xi \rightarrow 1^+$. Applying this limit to the Langrangian coordinate $\eta$, we obtain the following identity:
\begin{equation}
\lim_{\xi \rightarrow 1^+} \{\eta\} = \lim_{\xi \rightarrow 1^+}\left\{\left(\frac{\Omega}{2\upi}\right)^{1/2} \frac{\bar a^3}{3} (\xi-1)(\xi^2 + \xi + 1) \right\} = \left(\frac{\Omega}{2\upi}\right)^{1/2} \bar a^3 (\xi-1).
\end{equation}
Using this result, we may rewrite the solution for $c(\eta,\tau')$ in (\ref{eq:ch_fs}) as a function of our original variables, $c(\xi, \tau)$. This gives
\begin{equation} \label{eq:c_profile_fs}
c(\xi,\tau) = \varepsilon\exp\left\{-\bar a\sqrtsign{\Omega/2}(\xi-1)\right\}
\sin\left\{\Omega\tau - \bar a\sqrtsign{\Omega/2}(\xi-1)\right\},
\end{equation}
which in turn yields the following gradient profile:
\begin{equation} \label{eq:cgrad_profile_fs}
   \frac{\partial c }{\partial \xi}(\xi, \tau)
     = -\varepsilon \bar a\sqrt\Omega\sin\left(\Omega\tau - \bar a\sqrtsign{\Omega/2}(\xi-1) +\frac{\upi}{4}\right).
 \end{equation}
 Applying the limit $\xi \rightarrow 1^+$ alongside $\Omega \gg 1$ to (\ref{eq:c_profile2}) and (\ref{eq:cgrad_profile2}), those solutions reduce to expressions (\ref{eq:c_profile_fs}) and (\ref{eq:cgrad_profile_fs}) above. Notice that effect of the bubble curvature is lost ($\xi$ in the denominator of (\ref{eq:c_profile}) vanishes), consequence of the thin boundary layer approximation. Consequently, the steady-state (first) term contributing to the gradient in (\ref{eq:cgrad_wall2}), which dominates for small values of $\Omega$, is also lost.

\section{Pressure corrected radius}
\label{app:acorr}
We begin by rewriting the mass conservation equation (\ref{eq:mcons_phys}) without surface tension,
\begin{equation}
\frac{1}{a} \frac{\dd a}{\dd\tau} + \frac{1}{3p}\frac{\dd p}{\dd\tau}
=  \frac{\Lambda}{a^2p} \cgrad.
\end{equation}  
Integrating this equation in time results in
\begin{equation}
   a\: p^{1/3} = a_0\:\mathrm{exp}\left\{\int_{0}^{\tau} \frac{\Lambda}{a^2p}\cgrad \dd\tau'\right\} = \acorr
\end{equation}
where $\acorr(\tau)$ has been identified using (\ref{eq:acorr_def}).
Its time derivative is then
\begin{equation} \label{eq:dacorr}
   \frac{\dd \acorr}{\dd \tau} = \frac{\Lambda \acorr}{a^2p}\cgrad .
\end{equation}   
Since the prefactor multiplying the interfacial concentration gradient in (\ref {eq:dacorr}) is positive at all times, we can write
\begin{equation}
    \mathrm{sign}\left(\frac{\dd \acorr}{\dd \tau}\right) 
    = \mathrm{sign}\left(\left.\frac{\partial c}{\partial \xi}\right|_{\xi = 1} \right).
\end{equation}
It follows that the oscillating part of $\acorr$, namely $\acorr-\bar a$, is in phase with
\begin{equation}
\int^\tau_0 \frac{\partial c}{\partial \xi}(\xi=1,x) \:\dd x =  
-\frac{\varepsilon}{\Omega} \left[ 
\sin\left(\Omega\tau- \frac{\upi}{2} \right)
+\bar a \sqrt\Omega \sin\left(\Omega\tau- \frac{\upi}{4} \right) \right].
\end{equation}
Thus, equivalently, $\phi_\mathit{corr} = \phi_\mathit{grad} - \upi/2$.

\bibliography{he1-references}
\bibliographystyle{jfm}

\end{document}